\documentclass[10pt,a4paper,aps,PRD,twocolumn,showpacs,superscriptaddress]{revtex4-1} 
\usepackage[english]{babel}
\usepackage{url} 
\usepackage{hyperref}
\usepackage{amsmath}
\usepackage{slashed} 
\usepackage{graphicx}
\usepackage{enumitem}
\usepackage{float}

\usepackage{color}
\usepackage{xcolor}
\usepackage{bm} 
\usepackage{cleveref}
\usepackage[normalem]{ulem}
\usepackage{units}

\usepackage{multirow}
\usepackage{tabularx}

\usepackage{subcaption}

\begin{document}
\allowdisplaybreaks

\title{Differential Measurement of Trident Production in Strong Electromagnetic Fields}
\author{Christian F. Nielsen}
\affiliation{Department of Physics and Astronomy, Aarhus University, 8000 Aarhus, Denmark}
\author{Robert Holtzapple}
\affiliation{Department of Physics, California Polytechnic State University, San Luis Obispo, California 93407, USA}
\author{Mads M. Lund}
\affiliation{Department of Physics and Astronomy, Aarhus University, 8000 Aarhus, Denmark}
\author{Jeppe H. Surrow}
\affiliation{Department of Physics and Astronomy, Aarhus University, 8000 Aarhus, Denmark}
\author{Allan H. S\o rensen}
\affiliation{Department of Physics and Astronomy, Aarhus University, 8000 Aarhus, Denmark}
\author{Marc B. Sørensen}
\affiliation{Department of Physics and Astronomy, Aarhus University, 8000 Aarhus, Denmark}
\author{Ulrik I. Uggerh\o j}
\affiliation{Department of Physics and Astronomy, Aarhus University, 8000 Aarhus, Denmark}
\collaboration{CERN NA63}

\date{\today}

\begin{abstract} 
In this paper, we present experimental results and numerical simulations of trident production, $e^-\rightarrow e^-e^+e^-$, in a strong electromagnetic field. The experiment was conducted at CERN for the purpose of probing the strong-field parameter $\chi$ up to 2.4, using a 200 GeV electron beam penetrating a 400 $\mu$m thick germanium crystal oriented along the $\langle 110\rangle$ axis. 
For the current experimental parameters we found that the trident process is primarily a two-step process, and show remarkable agreement between theoretical predictions and experimental data. This paper is an extension of the previously published paper \cite{Niel_2023} and features new analysis differential in the energy of the produced positron and electron in the trident process. Even for the more demanding differential analysis, we find good agreement between theoretical predictions and experimental data, while a slight discrepancy is found in the high energy tail of the trident spectrum. This discrepancy could be an indication of the direct process, but further investigation is needed due to the large uncertainties in this part of the spectrum. Finally we present a suggestion for a future experiment, aiming to probe the direct process using thin crystals.

\end{abstract}

\maketitle
\section{Introduction}
When an electron impinges on an electrostatic potential barrier, it may penetrate or be reflected by the barrier. Classically, for electron energies less than the barrier height, the electron is always reflected. In non-relativistic quantum mechanics, an exponentially damped tunneling into the barrier is predicted, with no transmission far beyond the classical turning point when the potential remains higher than the electron energy. In relativistic quantum theory, however, an undamped electron-current is present beyond the classical turning point provided the barrier rises sufficiently abruptly and high, even if the barrier has an infinite height. This was shown in 1929 by Oscar Klein \cite{Klei29} for a step barrier in one of the first applications of the Dirac equation. 
It became known as the 'Klein paradox'.  As later shown by Fritz Sauter \cite{Saut31a,Saut31b}, inspired by a supposition by Niels Bohr, the potential has to rise with the rest energy of the electron, $mc^2$, over its reduced Compton wavelength, $\lambdabar_C=\hbar/mc$, for transmission to occur with substantial probability. The corresponding field strength,
\begin{equation}
\mathcal{E}_0 = m^2c^3/e\hbar \simeq1.32\times10^{16}~\mathrm{V/cm},
\label{eq:E0}
\end{equation}
later became known as the critical or Schwinger field.

Previous studies of the Klein paradox have been limited to theory \cite{Grei85,Krek04,Giac08}.Possibility of observing phenomena analogous to the Klein paradox in graphene have been reported \cite{Kats06,Calo06,Buch06,Boggild2017,Nguyen2018}. Other
studies have been partly motivated by heuristic arguments linking the Klein paradox, strong field pair production and Hawking radiation from black holes \cite{Mull77,Hols98,Hols99}. Today, the Klein paradox is explained by the creation of electron-positron pairs at the boundary \cite{Grei85}.

\begin{figure*}[ht!]
	\begin{center}
		\includegraphics[width=0.85\linewidth]{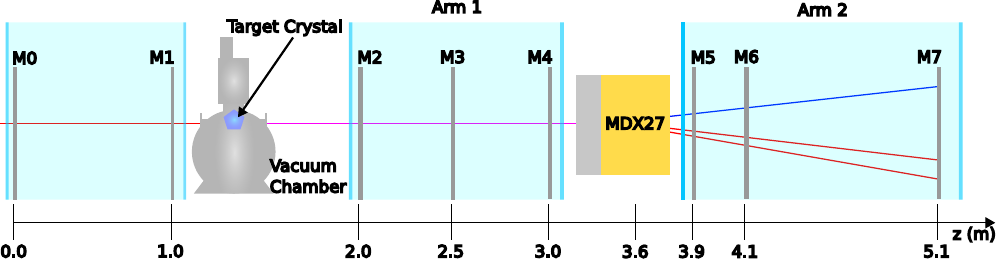}
\caption{Experimental setup. A schematic representation of the experimental setup in the H4 beam line in the SPS North Area at CERN. The symbols ``Mi'', with $i=0,\ldots,7$, denote `Mimosa-26' position sensitive detectors.}
		\label{fig:Setup}
	\end{center}
\end{figure*}

In this paper, which elaborates, underpins and extends the results presented in \cite{Niel_2023}, we present results for an analogous process, the trident production in strong electromagnetic fields. 
Based on a comparison of experimental values and simulated expectations computed using the Local Constant Field Approximation (LCFA), even when treated differentially in terms of energy of the produced positron, it is shown that the trident process $e^-\rightarrow e^-e^+e^-$ in a strong external field is well understood under our experimental conditions.  In \cite{Niel_2023} we only show the integrated positron spectrum, while this paper features differential spectra, which is a considerably stronger test of the underlying theory, as information is lost during integration.
Furthermore, details of the experiment and the data analysis are provided so the reader can verify the robustness of the results. 
Finally, we present simulated results for relevant cases which have not yet been investigated experimentally, mainly to inspire future experiments. These results are based on the theory thus corroborated by comparing the experimental values and the simulated expectations presented.

Charged particles exhibit strong-field effects when the strength of the external field in the instantaneous rest system approaches the critical field $\mathcal{E}_0$.
 One manifestation is that, due to the neglect of finite recoil, the classical synchrotron radiation spectrum for an electron in a constant magnetic field reaches angular frequencies higher than the electron's energy divided by $\hbar$, that is the classical spectrum contains photons which are more energetic than the electron itself. Quantum corrections remedy this situation, as has been shown in previous experiments \cite{PhysRevD.86.072001}. The parameter $\chi$, defined by \cite{Berestetskii_b_1989,ritus_1985,Baie98}
\begin{equation}
             \chi^2=(F_{\mu\nu}{p}^\nu)^2/{m^2c^2}\mathcal{E}_0^2,
 \label{eq:chi}
 \end{equation}
gives an indication of the importance of the strong-field effects. Here $F_{\mu\nu}$ is the electromagnetic field strength tensor and ${p}^\nu$ is the four-momentum of the particle. For a constant field perpendicular to the direction of motion, $\chi$ becomes exactly the strength of the field in the particle's system divided by $\mathcal{E}_0$ (the field is boosted by the Lorentz factor $\gamma$). Additionally, for a magnetic field, $\chi$ becomes equal to the ratio of $\hbar$ times the characteristic angular frequency of classical synchrotron radiation and the input energy \cite{Sorensen1996}, up to a numerical factor of order 1. Strong-field effects become significant for $\chi$ around 1 and larger. Hence, $\chi$ is known as the strong-field parameter and some literature refer to $\chi$ as "the quantum non-linearity parameter". For photons, the expression (\ref{eq:chi}) for $\chi$ applies as well, with the momentum of the electron replaced by the momentum of the photon.

The strong field is achieved in our experiment by aiming a beam of high-energy electrons at a single crystal, where a principal axis is aligned with the direction of the beam. Coherent scattering of electrons on rows of atoms, in this situation, implies that the electrons are effectively moving through the crystal as if they were subjected to the 'continuum' field achieved calculationally by smearing out the crystal atoms uniformly in the direction of the main axis. Hence, the motion, which determines radiation yields, trident production, and the like, is therefore determined by a field that, effectively, has a macroscopic extent in the axial direction. Conversely, if the major axial and planar directions of the crystal are far away from the beam direction, the atoms behave as if they were placed randomly and the radiation and trident processes will appear exactly as if the target were amorphous. Hence, we call this a 'random' orientation of the crystal, or, we designate it -- to be brief -- as an 'amorphous crystal'.

There are two contributions to the production of tridents. The first contribution occurs when incoming electrons produce electron-positron pairs directly in the continuum field or in the field from individual atoms. The second contribution occurs in two stages, where the electron first emits a photon, which then later converts to an electron-positron pair while passing the remaining part of the crystal. For the crystal used in this experiment, the two contributions are of the same order of magnitude for random orientation (corresponding to an amorphous target of the same thickness), while the two-step process is up to two orders of magnitude larger than the direct production in the aligned case. In the aligned case, any process resulting from coherent action of the crystal atoms is accompanied by an incoherent component due to thermal diffuse scattering on individual atoms. The latter usually gives only a small addition to the coherent contribution, except for the production of relatively low energy pairs in the direct trident process.

 \section{Experiment and Data Preprocessing}
 The experiment was performed by the NA63 collaboration at the H4 beamline of the CERN SPS that provided a 200 GeV electron beam having a $\sigma_x \simeq \sigma_y \simeq 105$ $\mu$rad divergence impinging on a 400 $\mu$m thick germanium single crystal oriented along the $\langle110\rangle$ axis. \Cref{fig:Setup} is a schematic of the setup where M0-M7 are MIMOSA-26 position sensitive CMOS-based pixel detectors \cite{Mimosa26}. The detectors have an active area of $1.1\times 2.1$ cm$^2$ containing $576\times1152$ pixels resulting in a resolution of a few $\mu$m (after weighting the hit pixels appropriately in the off-line analysis). The crystal target is mounted on a goniometer that allows us to set the crystal orientation, aligned or random, with $\mu$rad precision. The MDX27 magnet provides an integrated magnetic field of $0.072$ Tm and, together with the detectors in Arm 1 and Arm 2, forms a magnetic spectrometer used to measure the energy of each charged particle from the deflection angle in the magnet. The crystal is situated inside a vacuum chamber at $\simeq 300$ K. To reduce scattering and background, all mimosas are placed in closed compartments that are continuously flushed with helium. 
The total material contributing to the background before the MDX27 magnet, in units of the radiation lengths, amounts to $\simeq 1.1\%$.

\subsection{Alignment of detectors} \label{sec:AlignmentOfDetectors}
Data from Mimosa detectors consist of a list of $(x,y)$ positions from clusters of pixels on each chip that have been recognized as a hit by the Mimosa preprocessing software. These $(x,y)$ hit coordinates are defined with respect to the coordinate system of each Mimosa, upon which we transform each Mimosa hit coordinates into a common coordinate system. This is achieved by letting M0 and M1 define the common coordinate system and employing an alignment algorithm that transforms hit coordinates into the common coordinate system. In practice this is done by an affine transformation that involves multiplying each hit by the $3\times3$ transformation matrix $A$ defined by
\begin{equation}
    X_i = A_i X_i'\label{eq:AlignmentTransformation}
\end{equation}
where 
\begin{equation}
    X_i = \begin{pmatrix}
    x_{i1} & x_{i2} & & x_{in}\\
    y_{i1} & y_{i2} & ...&y_{in}\\
    1 &1 & &1
    \end{pmatrix}, \end{equation}
    and
    \begin{equation}
    X_i' =     
    \begin{pmatrix}
    x_{i1}' & x_{i2}' & & x_{in}'\\
    y_{i1}' & y_{i2}' & ...&y_{in}'\\
    1 &1 & &1
  \end{pmatrix},
\end{equation}
which is the hits in the common coordinate system and in the Mimosa $i$'s coordinate system respectively. This transformation provides an alignment matrix for each detector after M1 and M2. This is an iterative process in which we slowly transform a Mimosa into a common coordinate system.

\begin{figure*}[ht!]
	\begin{center}
		\includegraphics[width=1\linewidth]{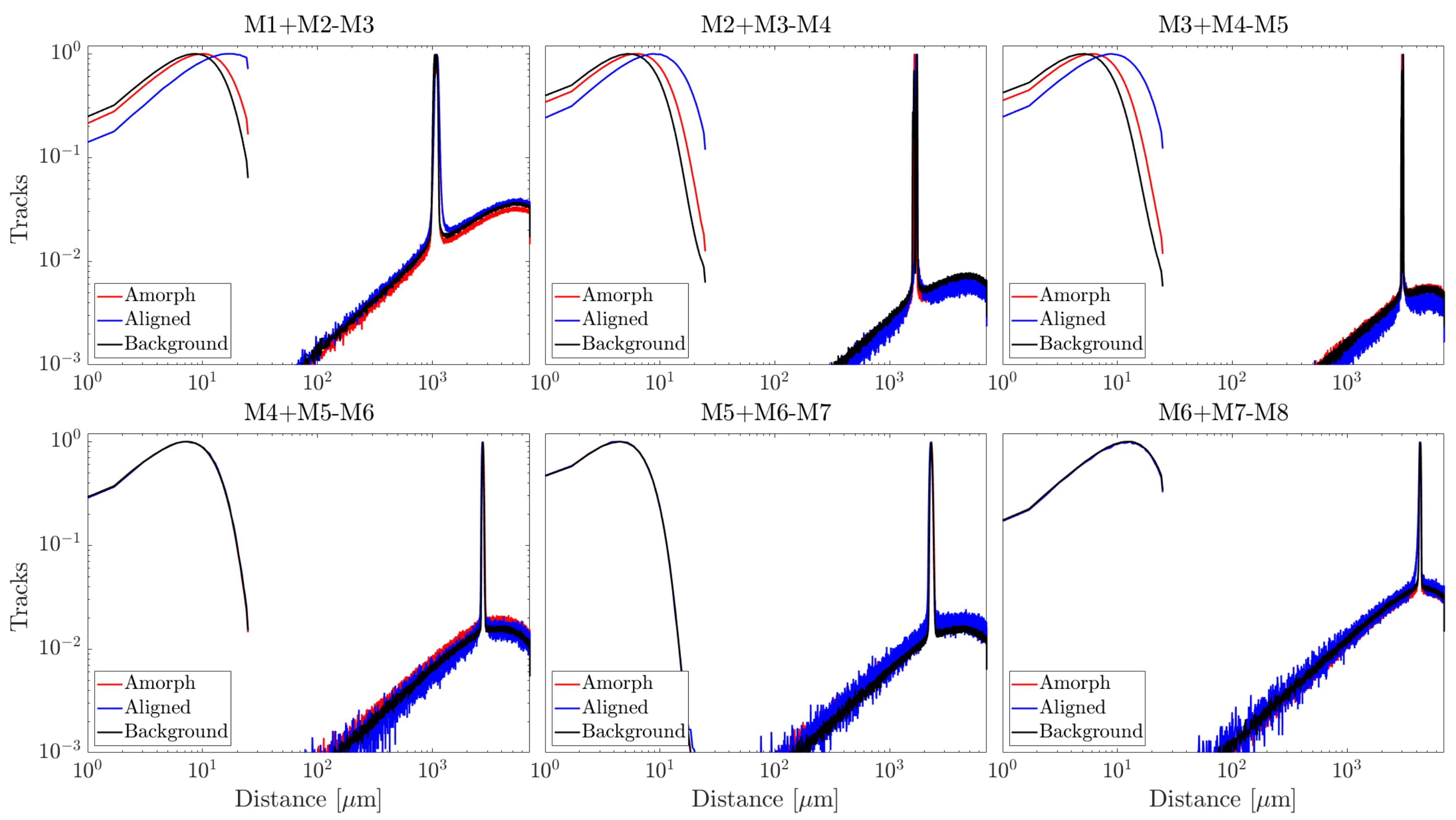}
            \caption{The distance between projected hits and actual hits in the Mimosa detectors, before and after using the detector alignment algorithm. Blue is for the aligned crystal data, red is for the amorphous data, and black is data without the crystal. Each crystal configuration is shown twice on each figure, one before alignment (right) and one after (left).
            Each figure corresponds to a particular detector, where the previous two detectors has been used to define a projected hit in the target detector, indicated by the title of each figure.    }
		\label{fig:DistanceBetweenActualAndExpectedAlignment}
	\end{center}
\end{figure*}

 The alignment algorithm begins by finding the first iterations lists $X_i$ and $X_i'$, for a considerable amount of events, by producing a set of tracks derived by combining all hits in $M_{i-2}$ and $M_{i-1}$, which are projected onto $M_i$ for each event. Around each track projection on $M_i$ we identify the closest hit within a search radius $R$. If we find a hit, we save the projected hit in $X_i$ and the actual hit in $X_i'$. We then find the corresponding matrix $A_i$ by solving \cref{eq:AlignmentTransformation} and transform all hits in $M_i$ using the matrix $A_i$. 
 This procedure moves $M_i$ in the common coordinate system, which could result in hits that were previously not within the original search radius, to now be included, and vice versa. A new list of projected and actual hits, $X_i$ and $X_i'$, with the same radius $R$ are found, and a new matrix $A_i$ is found, and the process of transforming all hits repeats. This process is continued until the matrix $A_i$ no longer changes significantly, after which the search radius is halved and the entire process is repeated. The search radius is lowered until the minimum radius $R_\mathrm{min} = 25$ $\mu$m is reached. At this point, the complete transformation for the target Mimosa is then the product of all the transformations done during this process, resulting in a final transformation matrix $A_i$. 
 After aligning $M_i$, we can use the newly transformed hits in $M_i$ together with $M_{i-1}$ to align the next Mimosa $M_{i+1}$.

\begin{figure}[ht!]
	\begin{center}
		\includegraphics[width=\linewidth]{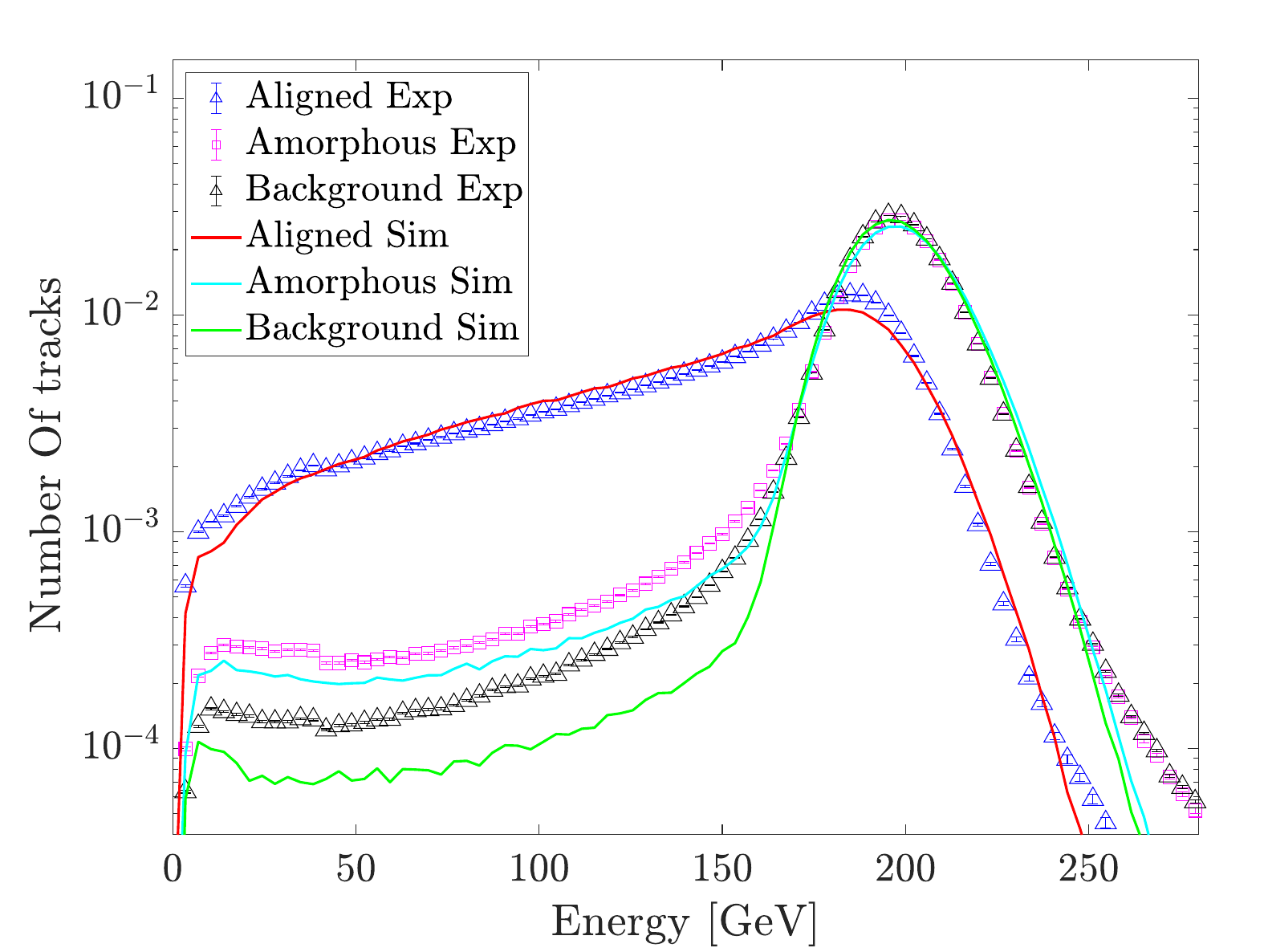}
            \caption{Simulated and experimental spectra for all primary particles for the aligned, amorphous and background configurations. }
		\label{fig:ElectronSpectrum}
	\end{center}
\end{figure}

In \cref{fig:DistanceBetweenActualAndExpectedAlignment}, we show the distance between projected and actual hits before and after the alignment process, for Mimosas M2-M7. The right hand side curves on each figure are the distances before alignment for amorphous, aligned and background (when the crystal is removed) curves, where the curves on the left of each figure are the same data, but after alignment. Before alignment, it is clear that noise hits and hits not pertaining to the particle tracks are contributing, but there is also a large peak originating from real particle tracks. This indicates that the physical alignment of the detectors in the laboratory is off by several mm. After alignment, we see that the error between the actual and projected hits are less than 20 $\mu$m. The alignment error is a combination of the detector uncertainty, which is about $3.5$ $\mu$m, and the multiple Coulomb scattering each particle undergoes between each detector.
During the two-week experiment, the detector's location drifted by more than 100 $\mu$m due to day to night temperature fluctuations, thus rendering the single alignment runs unusable. The alignment method predicts where a particle is likely to be in a detector. This is very easy for a freely moving particle with no magnets. The alignment algorithm was therefore adjusted to be able to align M5-M7 with the magnet turned on. The adjusted algorithm produces an alignment matrix for each set of $10^5$ events, so that the aligned, amorphous and background runs are divided into sets of $10^5$ events that are individually aligned. By assuming that the energy of each particle is known, we can determine the deflection in the magnet, allowing us to predict the track in detectors M5-M7. For amorphous and background configurations, the material budget is only a few $\%$ of the radiation length $X_0$, meaning that most particles will have the original 200 GeV. In \cref{fig:ElectronSpectrum}, we show the simulated and experimental data of the primary electron spectrum for all three crystal configurations.
In the amorphous and background configurations, it is clear that most particles lose no energy and have an energy of 200 GeV when going through the magnet. The energy peak in the aligned configuration is $185$ GeV, which is the energy we assume all particles have going through the magnet in the aligned configuration. To avoid using low energy particles in the alignment algorithm, we confine the algorithm to only use tracks that are within $\pm$ 20 GeV of the peaks of each configuration. This peak is 200 GeV for the amorphous and background configurations and 185 GeV for the aligned configuration.

The difference between the amorphous, aligned and background curves in \cref{fig:DistanceBetweenActualAndExpectedAlignment} after alignment of the detectors is due to additional scattering in the crystal. As will be discussed later, the particles scatter more in the aligned crystal configuration compared to in the amorphous crystal configuration. 
We believe the discrepancies between the simulated and experimental data in \cref{fig:ElectronSpectrum} for the background and amorphous configurations is due to a small number of particles scraping the collimators in the beamline, which is not accounted for in the simulation. This effect is largest for the background case and is negligible for the aligned case because the material budget in the setup, which we account for in the simulation, is much larger than what is scraped in the beamline collimators.

Since the energy distributions of particles are not perfectly Gaussian, the alignment algorithm for the detectors after the magnet could introduce a slight bias. To account for this, mimosas M6-M8 are aligned using the same procedure in the simulation, therefore introducing an identical bias in the simulated data. This improved alignment procedure in the simulation is new, compared to the results shown in \cite{Niel_2023}, and results in minor differences in the simulated curves that will be mentioned when relevant. The alignment procedure for the experimental data have not changed compared to \cite{Niel_2023}.
 
\subsection{Alignment of Crystal Target}
As the penetrating particle approaches major crystallographic directions, like low-index planes and axes, its radiation emission is typically enhanced by factors of 4-5 in the planar case, and up to 60 for the axial case, for a Ge crystal with the thickness used in this experiment \cite{MEDENWALDT1990517}. Thus, a measurement of the radiation emission, as a function of angular setting of the crystal, identifies the planes of the crystal by the construction of a stereogram. The location, in angular space, of the axis is found from the stereogram and verified by a couple of scans across the axis,  which must be symmetric if the correct location is found. For this experiment the crystal was mounted on a goniometer with a stepsize of $1.7~\mu$rad.  This crystal alignment technique provides a $\sim10~\mu$rad positioning precision on the $\langle110\rangle$ axis.

\begin{figure}[ht!]
	\begin{center}
		\includegraphics[width=\linewidth]{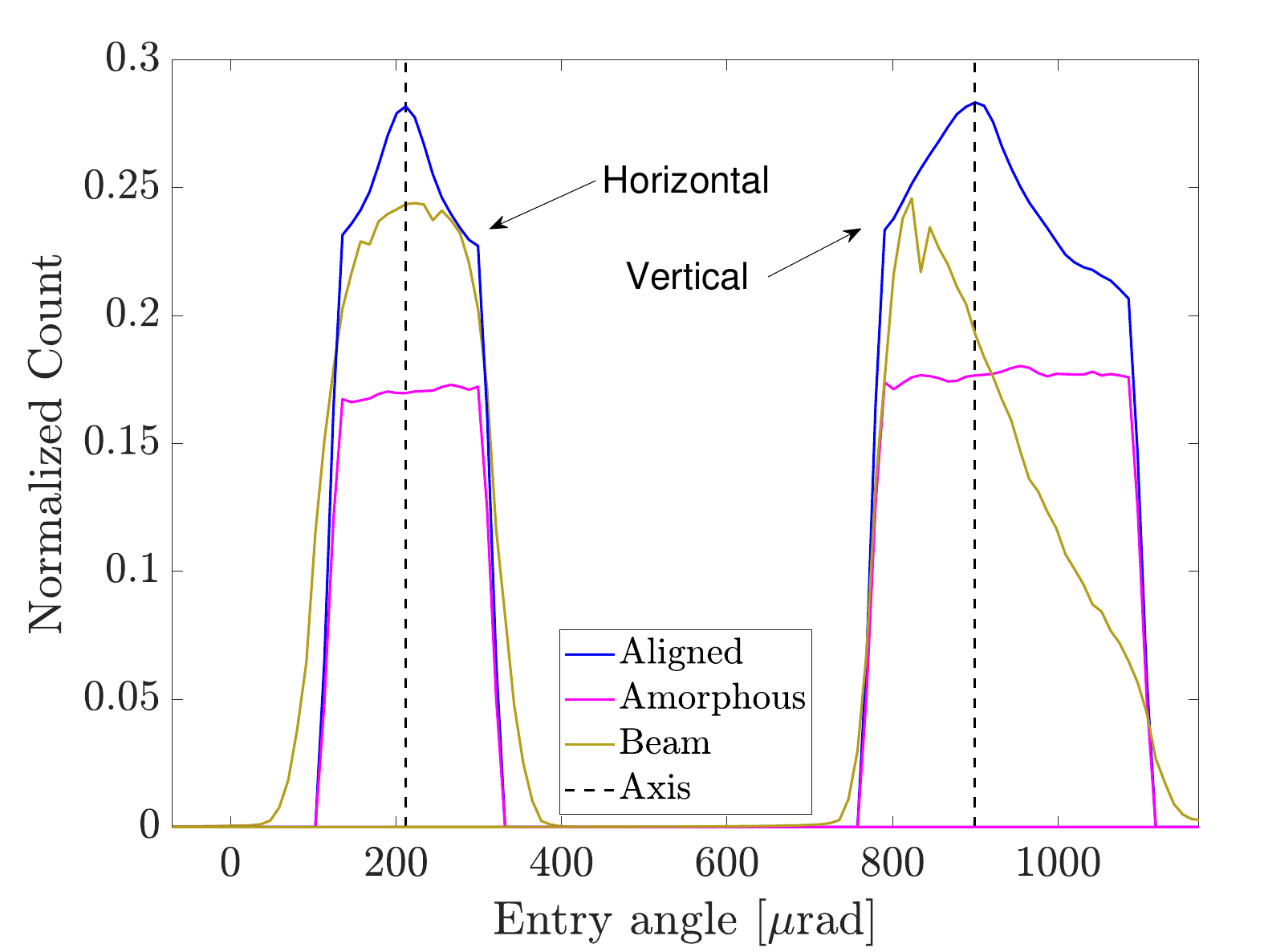}
            \caption{Amount of primary electrons losing more than 90 GeV as a function of vertical and horizontal entry angle into the setup, normalized by beam shape for the amorphous and aligned crystal orientations. The beam curve correspond to the full beam normalized to fit the scale of the other curves. Black vertical lines indicate the axis location in angle for the aligned case.}
		\label{fig:OfflineAxisScan}
	\end{center}
\end{figure}
In the off-line analysis, the angular location of the axis is identified by the impinging particle losing a substantial fraction of its energy, through radiation emission, its primary process for energy loss.
In \cref{fig:OfflineAxisScan}, we show the amount of particles losing more than 90 GeV as a function of the horizontal ($x$-direction) and vertical ($y$-direction) entry angle defined by hits in M0 and M1, normalized to the full beam. Since there are no directional effects in the amorphous crystal orientation, which was verified by small angular scans around the chosen 'amorphous' location, this curve is flat. Clear peaks indicate the direction of the axis in the horizontal and vertical planes for the setup in the aligned crystal configuration. The beam is highly collimated which is what gives rise to these sharp edges in the beam profile. In the simulation of the experiment, which will be described next, we need to take into account the beam shape and axis location from the offline analysis. 

\section{Simulation}
The simulation tool is built using two separate codes, one simulating the experiment, and the other simulating the particles penetrating the target crystal. 
Since the material budget for each element is small, the time steps for propagating particles through the setup can be macroscopic, whereas for aligned crystals the time steps need to be on the nanometer scale.
Thus, a simulation tool was developed for the experiment while an extended version of the crystal simulation used and explained in \cite{NIELSEN2022,Nielsen_2020,CFN2019GPU,NJP2021} was included in the experimental simulation.
In the following, we will denote the entire simulation which includes both experiment and crystal code "simulation", whereas the results based solely on the crystal simulation will be called "crystal simulation". 

The purpose of the simulation is to produce a datafile identical to the ones obtained from the experiment. The result of a simulation is therefore a list of events containing $(x,y,z)$ positions from hits in detectors as in the experiment.
Each event contains a primary electron with energy of 200 GeV, whose starting position and entry angle matches that of the experimental beam. A list of initial conditions is generated from an experimental data file based on accepted single particle tracks. A random entry in this list is chosen for each particle in each event, defining the particle's initial conditions. The particle is then propagated through all the elements in the setup, including mylar windows, tape, helium, air detectors, and the crystal. With the exception of the crystal, all elements are divided into 10 pieces through which the particle propagates through freely. After each section of material, random numbers are drawn and compared to probabilities of photon emission, pair production, trident production and multiple Coulomb scattering. If any event occurs, a value is picked from the event's underlying distribution through inverse transform sampling as described later; when, for example, a photon with energy $E_\gamma$ and opening angle $1/\gamma$ is emitted, the particle loses $E_\gamma$ and continues in its original direction. The secondary particle is now propagated to the end of the experiment in the same manner as the initial particle. 

If a charged particle encounters a detector, the detector will record a hit with a probability of approximately one. Both the $x$ and $y$ positions of the impact coordinate are added with a random number derived from a normal distribution with mean zero and width $\sigma = 4.2$ $\mu$m.
 In the event that a secondary particle impacts within $50$ $\mu$m of an already recorded hit, the two hits will be combined into a single hit with their average position.  This is intended to simulate how a detector in such a situation would react.
  When a charged particle strikes the surface of the detector chip, a cluster of pixels are triggered, leaving several pixels in the vicinity of the impact active. Pixels are about 20 $\mu$m wide, so if two particles land close to each other, their active pixel areas will overlap. The preprocessing software in the detectors is more sophisticated than simply combining two adjacent hits as mentioned above, but we consider this implementation to be adequate. The detectors for this experiment do not record photons.

The detectors have $x$ and $y$ dimensions of 2.1 cm and 1.1 cm, respectively, and are positioned according to the alignment of the detectors during the experiment. The dipole deflection occurs in the horizontal direction ($x$), where the detector is the widest. The finite dimension of the detectors also implies that if a particle does not land within the boundaries of the detector, the hit will not be recorded. This leads to a natural energy cutoff for low energy particles, since they are simply deflected outside detectors M6-M7. It is for this reason that M6-M7 are placed so close to each other and the magnet, whereas M8 is located a considerable distance away in order to increase the energy resolution for high-energy particles. For the simulation, is important that the direction of the beam, position of the detectors in space, and absolute direction of the crystal axis relative to the beam, match what is measured in the experiment, as all three parameters affect one another significantly. 

Since particles in the GeV range experience only small-angle deflections in the CERN supplied MDX dipole magnet we can safely approximate the magnet as an instantaneous deflection only at its center. While this is computationally easier, it also excludes electromagnetic processes arising from the interaction with the magnetic fields. Nevertheless, this neglect is of no concern since possible photon emissions will be in the keV range, which eliminates any possibility of pair production. 

The crystal simulation is initiated when a particle penetrates the target crystal. The Boris Pusher algorithm is used to propagate charged particles in the external electric field \cite{Ripperda_2018_Boris,BorisPusher,BorisPusher1970}. The photons travel in a straight line through the field without any perturbations in their direction of motion. With each step forward in time, we determine if any electromagnetic process occurred, and if so, we determine the properties of the event based on its underlying distribution. 
A detailed description of the exact methodology for evaluating random numbers from distributions is provided in \cite{NIELSEN2022}, which describes how inverse transform sampling is used in the case of photon emission and pair production using the LCFA. 
The relevant theoretical models are described later in this section as well.

When a real photon is produced, it is emitted at a $1/\gamma$ angle, with a random azimuthal angle between 0 and $2\pi$, relative to the emitting particle. The emitting particle receives an instantaneous recoil, causing it to slow down. When a pair is created from a direct trident process, the emitting particle also receives an instantaneous recoil according to the total energy of the pair, while the pair is separated in opposite directions according to the Borsellino angle \cite{borsellinoPairAngle}, with a random azimuthal angle between 0 and $2\pi$. When a photon decays, the photon simulation stops, and the produced electron and positron receive a transverse kick in opposite directions according to the Borsellino angle at random azimuthal directions.

\begin{figure*}[ht!]
	\begin{center}
		\includegraphics[width=0.7\linewidth]{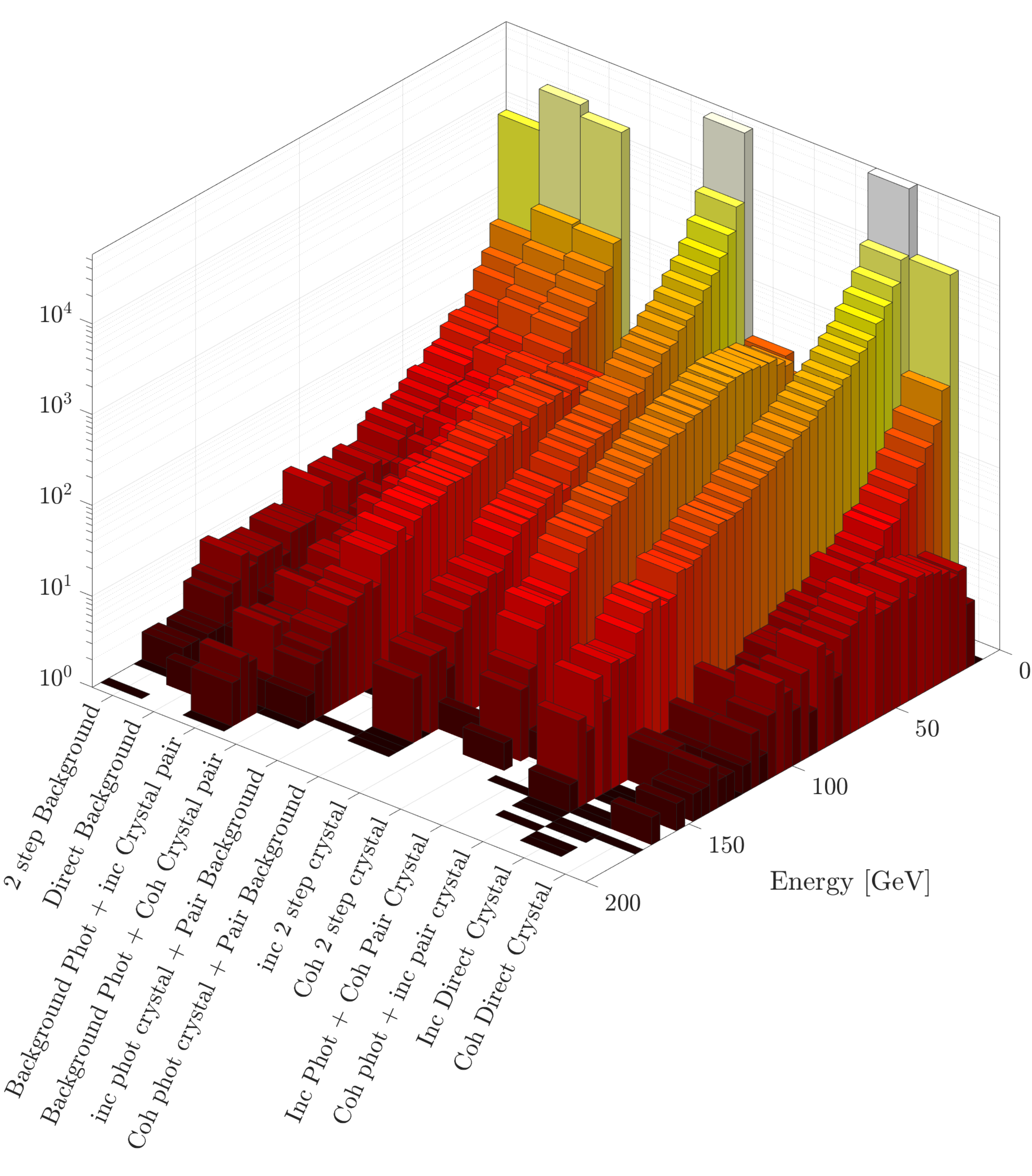}
\caption{ Simulated positron spectra categorized by their type of origin in the experiment. A description of the different origins can be found in the text. All processes happening in the coherent electric field from the crystal, is labeled "coh" (coherent), while the incoherent scattering processes are labeled "inc" (incoherent).}
		\label{fig:TridentTypes}
	\end{center}
\end{figure*}

\begin{figure*}[ht!]
	\begin{center}
		\includegraphics[width=0.7\linewidth]{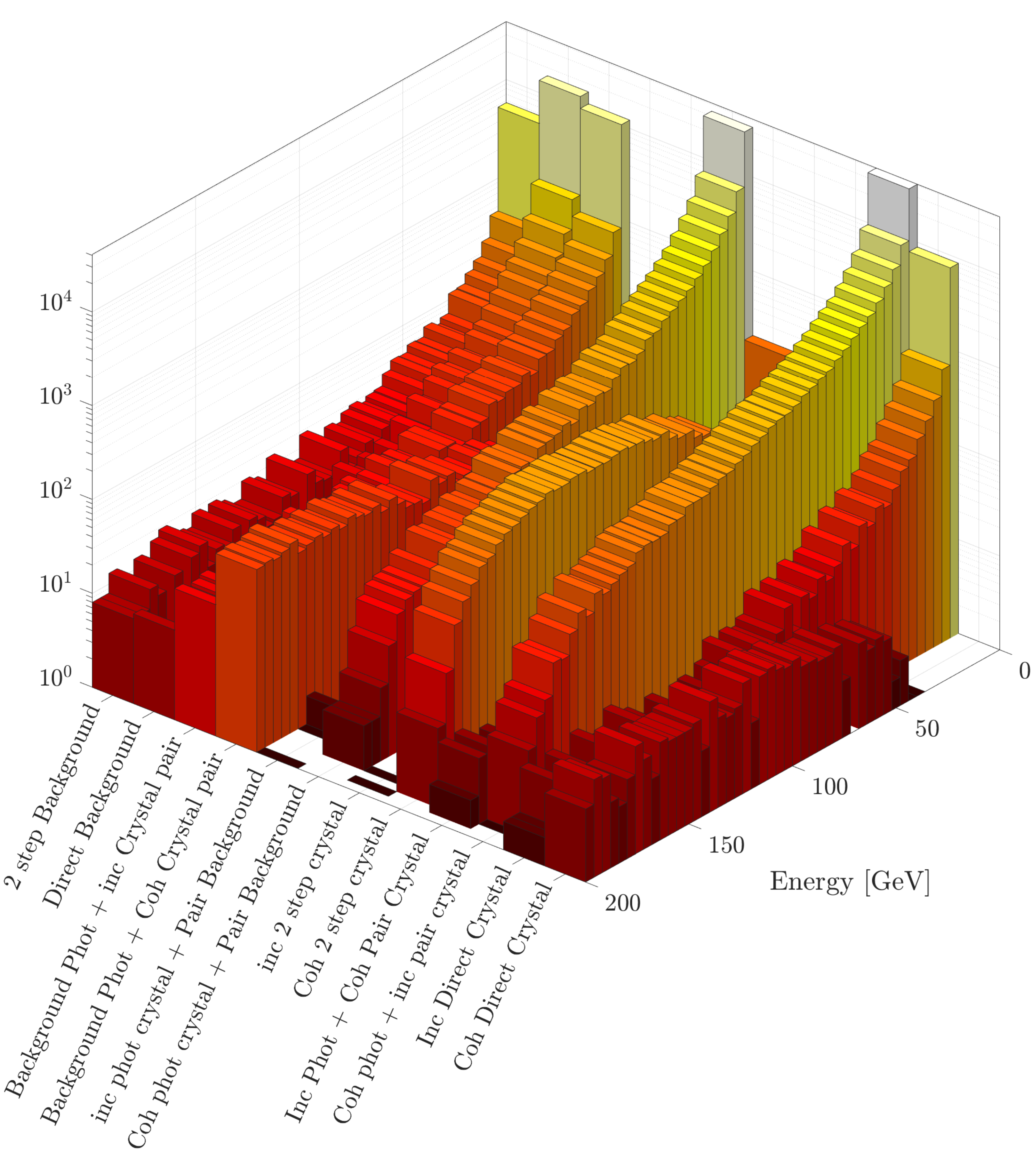}
            \caption{  Simulated pair energy spectra categorized by their type of origin in the experiment. A description of the different origins can be found in the text. All processes happening in the coherent electric field from the crystal, is labeled "coh" (coherent), while the incoherent scattering processes are labeled "inc" (incoherent).}
		\label{fig:TridentTypesPhotonEnergy}
	\end{center}
\end{figure*}

On \cref{fig:TridentTypes} and \cref{fig:TridentTypesPhotonEnergy} we show the positron spectrum and pair energy spectrum for all tridents produced in the simulation. The pair energy is defined as the sum of the electron and positron energies produced during the reaction. These spectra are categorized by the various ways a trident can originate.
It is instructive to note that the number of ways in which electrons and positrons can be created is greater than the fundamental number of trident processes that will be discussed in the following sections.
In addition to the usual two-step and direct processes, there are cross processes, such as the creation of a photon in the background, the production of that pair in other background elements, or the production of that pair through incoherent and coherent processes within the crystal.

Additionally, a photon can also be generated in the crystal by coherent or incoherent processes, and pair produce in the subsequent background elements located after the crystal. While the experimental setup cannot distinguish between these processes, it is clear that the pure crystal processes will dominate the spectrum. 
As we also see, all incoherent pair-production processes diverge at low energies, where coherent pair-production processes decline as a result of the small value of the strong-field parameter $\chi$, \cref{eq:chi}, for low-energy photons, resulting in an exponentially suppressed probability of pair-production.
As expected, we see a very low contribution from the coherent direct trident process. 
At the end of this paper, a brief discussion is provided on how to enhance the visibility of the direct trident process in crystals.

\subsection{Crystal Fields}
The motion of a charged particle incident at a small angle to a major crystallographic direction is in first approximation governed by successive, correlated small-angle collisions with screened target nuclei. The trajectory of the particle is determined by the continuum potential obtained by smearing the atomic charges along the axis with which it is nearly aligned, \cite{Lind65, JUAnotes} and \cite{Sorensen1996, Ugge05}. For a row of atoms the continuum potential is given by
\begin{equation}
   U(r)\; = \; \frac{1}{d}\int_{-\infty}^{\infty}\text{d}zV(r,z) 
               \; ,    \label{eq:contPot}
\end{equation}
where $V$ refers to the potential energy pertaining to the interaction between the projectile and a target atom, $z$ is the coordinate along the atomic row, $r$ is the transverse distance to the center of the axis, and $d$ is the average spacing between the atoms in the string. For a single isolated row of atoms, $U$ has rotational symmetry, Eq.\ (\ref{eq:contPot}). For a true crystal there is periodicity in transverse space, $U=U({\bf{r}})$. In the crystal simulation, we model the electric field using the Doyle-Turner potential which is based on an analytical approximation to relativistic Hartree-Fock atomic potentials. For a single row of atoms and a projectile with unit charge, the potential is given by
\begin{equation}
   U(r)\; = \; \pm\frac{e^2}{a_0}\frac{2a_0^2}{d}
        \sum_{i=1}^4\frac{a_i}{C_i} 
         \, e^{-r^2/C_i} 
               \; ,    \label{eq:DT}
\end{equation}
where the sign reflects that of the incoming charge ($\pm \vert e\vert$), $a_0=0.5292$ {\AA} is the Bohr-radius of hydrogen,
\begin{equation}
  C_i=C_i(\rho )\equiv b_i/4\pi^2+\rho^2
            \;  ,       \label{eq:DTCi}
\end{equation}
and $\rho$ denotes the two-dimensional root-mean-square thermal displacement of the atom from its equilibrium position. For details and explicit values of the coefficients $a_i$ (\AA ) and $b_i$ (\AA$^2$), see \cite{Doyl68} and \cite{Ande82}. For values of the thermal vibration amplitude $\rho$, see \cite{Nielsen_1980}. The Bohr radius is proportional to the reduced Compton wavelength of the electron by $a_0=\lambdabar_C/\alpha$, where $\alpha$ is the fine structure constant.

Due to the thermal vibrations of atomic nuclei, there is a local density of nuclei surrounding each string of atoms that is given by
\begin{equation}
\label{eq:nnuclstring}
    n_n(r)=\frac{1}{\pi\rho^2d}e^{-r^2/\rho^2}.
\end{equation}
It is this density distribution that is combined with \cref{eq:contPot} in order to include the thermal vibrations of the atomic nuclei in the resulting potential, \cref{eq:DT}. Since the string potential is highly dependent on the thermal vibration amplitude, elements in which the thermal amplitude is sensitive to the crystal temperature can be used to probe several values of the strong-field parameter $\chi$. Germanium, for example, can achieve almost a factor of two in the maximum field strength by cooling the crystal from 300 K to 100 K. In comparison, the effect of cooling on tungsten is negligible.

\begin{figure}[ht]
\centering
		\includegraphics[width=\linewidth]{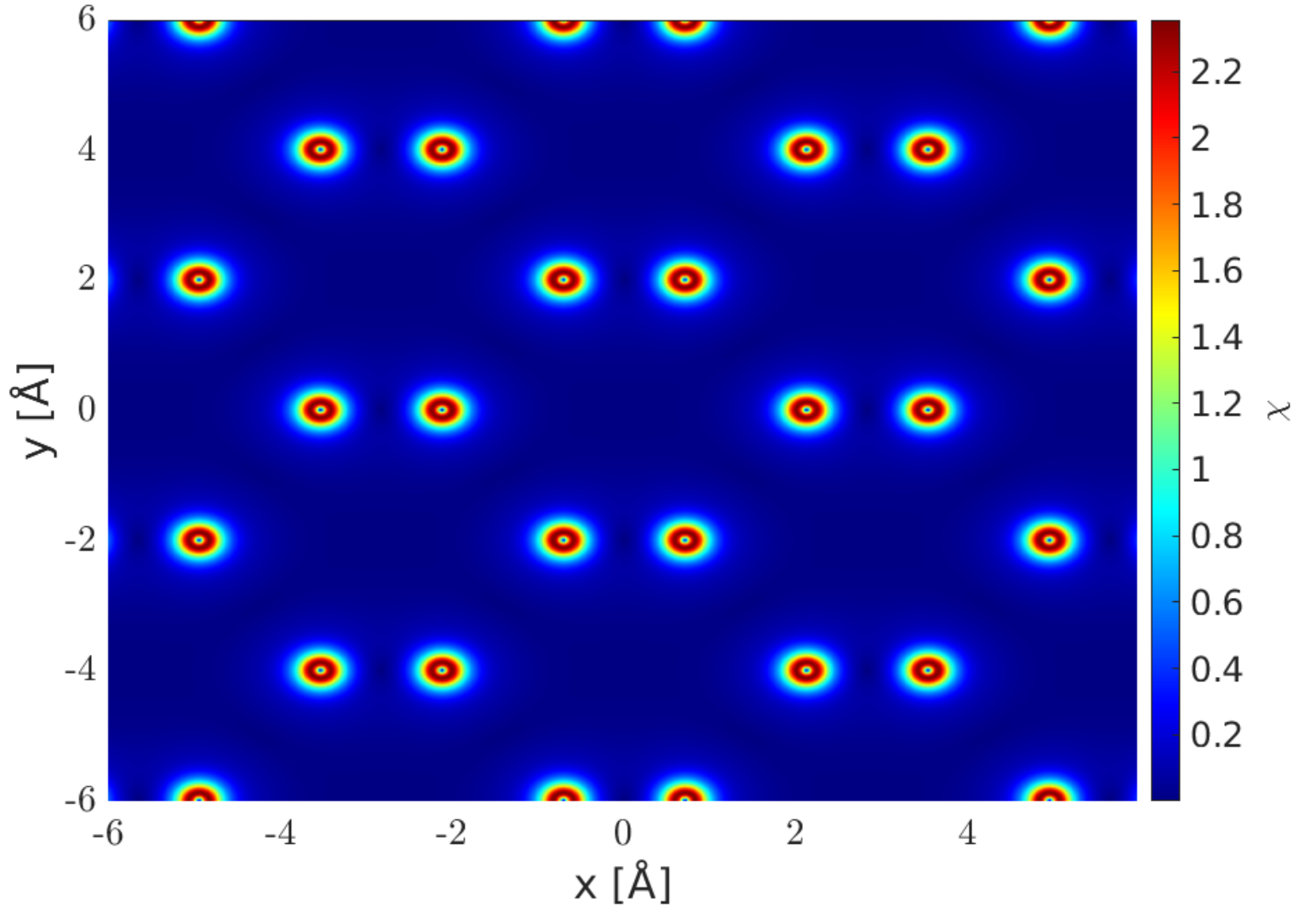} 
\caption{The strong-field parameter $\chi$ inside a germanium crystal kept at room temperature and oriented along the $\langle 110\rangle$ axis for a 200 GeV electron. }
\label{fig:ElectricFields}
\end{figure}

During each step in the crystal simulation, a particle experiences only the field originating from the 20 closest strings. A transverse cut-out of the $\chi$-values encountered for a 200 GeV electron incident on a germanium crystal oriented along the $\langle 110\rangle$ axis is shown in Figure \ref{fig:ElectricFields}. Each peak represents a string of atoms that extends into the figure for the entire length of the crystal, and the intensity axis displays the resulting strong-field parameter $\chi$, \cref{eq:chi}, assuming that the electron has zero transverse momentum. 
Since the field from a single string points radially away from it, the total field, and hence $\chi$, will be zero if a particle is placed between neighboring strings symmetrically. For this particular crystal and electron energy, we are able to probe $\chi$ values in the range $0 < \chi < 2.4$. 

For a thorough discussion of the motion of charged particles in aligned single crystals, the reader should refer to the original publication of J. Lindhard \cite{Lind65}, the extensive lecture notes of J. U. Andersen \cite{JUAnotes}, as well as review articles: \cite{Sorensen1996, Ugge05}.

In the case of a negatively charged particle, such as an electron, each string acts as a potential well, which can confine the particle's motion transversely. This is an example of so-called channeling. Using the potential depth and the particle's Lorentz factor $\gamma$, we can determine what is called the ``critical angle'' or ``Lindhard angle''. Generally, this angle represents the incidence angle to the crystal axis below which a high fraction of incoming particles will be channeled, because the energy associated with their transverse motion is initially lower than the string potential maximum (zero for electrons). For the axial case, where it is usually denoted $\psi_1$, the critical angle is
\begin{equation}
 \psi_1\; = \; \sqrt{ \frac{4Z e^2}{pvd}}\; = \;  \frac{\alpha}{\sqrt{\gamma}\beta}\sqrt{\frac{4Za_0}{d}}
       \;  \label{eq:psi1}
\end{equation}
for a unit-charge impact at momentum $p=\gamma mv$ and velocity $v=\beta c$ on a target of atomic number $Z$. It should be noted that $\psi_1$ scales as $1/(pv)^{1/2}$, that is, for high values of the Lorentz factor, it decreases in proportion to $1/\sqrt{\gamma}$ where $\gamma\equiv 1/\sqrt{1-\beta^2}$. In the case of a 200 GeV electron incident on a germanium crystal oriented along the $\langle 110\rangle$ axis, the critical angle is $\psi_1 = 57$ $\mu$rad. 

\subsection{Electromagnetic Processes in External Fields}\label{sec:EMProcesses}
For the crystal simulation, we have implemented three processes in the external continuum field: Photon emission, pair production, and direct trident production. The same processes are implemented for incoherent scattering on nuclei, which is explained in detail in Subsec. \ref{sec:AMProcesses}.  

Photon emission and pair production are modeled using the local constant-field approximation (LCFA). 
The applicability of the LCFA for photon emission requires the projectile to move only slightly in the transverse direction relative to the strings of atoms during the formation of the radiation. Because coherent photon emission is possible from points along the trajectory that are covered by the same light cone, it is necessary that the opening angle of the light cone, $1/\gamma = 2.6$ $\mu$rad, be smaller than the angular excursions of the projectile during passage through the crystal.
For a primary electron in the channeling regime, its angular excursions is on the order of the critical Lindhard angle \cite{Lind65,JUAnotes}, which in our case is  57 $ \mu$rad. For entry angles  greater than the critical Lindhard angle, but smaller than the Baier angle $U_0 / mc^2 = 0.4 $ mrad, the angular deflections remain larger than $1/\gamma$ \cite{Sorensen1996}. Here, $U_0$ is the continuum string potential depth which amounts to 215 eV for a single $\langle110\rangle$ row of Ge at room temperature. The LCFA is therefore appropriate for nearly all particles in a beam with a divergence of $\simeq 105$ $\mu$rad aimed at the aligned crystal.
See, e.g. \cite{cern2017,NielsenLCFA2022}, for studies demonstrating the applicability of the LCFA in crystals and short focused laser pulses.
As a  measure of the applicability of the constant-field approximation under channeling conditions, the authors of \cite{Baie98} introduced the parameter $\rho_c = \xi^2 = 2U_0\gamma/mc^2 $  (where $\xi$ is known as the classical non-linearity parameter \cite{AntoninoReview,FedotovReview2022} in the strong-field laser community), which is $\simeq 330$ for 200 GeV electrons under channeling conditions. The large value of $\rho_c$ verifies that treating the local field as constant is a good approximation for the photon emission in the two-step process.

The requirements for application of the LCFA in the pair creation vertex are similar by crossing symmetry. However, replacement of the primary electron energy by the lower photon energy makes conditions less favorable.
Yet this is a relatively minor concern, since the coherent pair-production rate for Ge$\langle 110\rangle$ is only higher than the incoherent pair-production rate if the photon energy is a larger fraction of the primary electron energy: averaging over transverse position at room temperature, the two rates are equal around 90 GeV. 

For the two-step process, the theoretical model described in this section averages over the photon polarization. For axially aligned crystals, the experimentally measured photon spectrum is polarization averaged since each projectile has a unique trajectory through the crystal. 
This is due to the unique angle and position of entry for each particle, and to multiple scattering altering each trajectory at random.
Therefore, due to the uniqueness of the emitting particle, a real photon emitted with a specific polarization will also follow a unique trajectory. Consequently, the pair production process also becomes polarization averaged, and modeling the real photon as unpolarized is a good approximation. 

The probability per unit time for an electron to emit an unpolarized photon in a locally constant strong electromagnetic field is given by \cite{Baie98,ritus_1985,Berestetskii_b_1989,baier1968quasiclassical}
 \begin{equation}\label{PhotonProb}
\frac{\text{d}P_{\text{rad}}}{\text{d}t} = -\alpha \frac{c}{\gamma\lambdabar_C} \int_0^{\infty} \text{d}u \,  \frac{5u^2 + 7u + 5}{3(1+u)^3 z} \text{Ai}'(z),
 \end{equation}
 where $\text{Ai}'(z)$ is the derivative of the Airy function $\text{Ai}(z)$ \cite{abramowitz+stegun}, $z = [u/\chi]^{2/3}$, $E$ is the electron energy. The factor in front of the integral may alternatively be expressed simply as $c/\gamma a_0$ or as $\alpha e\mathcal{E}_0c/E$, where $\mathcal{E}_0$ is the critical field, \cref{eq:E0}. In each time step, we evaluate the absolute probability of emission, and if a photon is emitted, we draw a random number based on the distribution of the photon energy spectrum, which is given by  \cite{ritus_1985,Berestetskii_b_1989,baier1968quasiclassical}
 \begin{multline}
 \frac{\text{d}P_{\text{rad}}}{\text{d}x\text{d}t}  = - \alpha \frac{c}{\gamma\lambdabar_C} \left\{\int_z^\infty \text{d}t \text{Ai}(t) \right.\\  + \left.\frac{\text{Ai}'(z)}{z} \left[ 2 + \frac{x^2}{(1-x)}\right] \right\}.
\label{PhotonSpectrum}
\end{multline}
The above equation is expressed in terms of the the ratio $x = E_\gamma / E$ which relates to $u$ by $u = x/(1-x)$ where $E_\gamma$ is the emitted photon energy.

Pair production from photons is treated in a similar manner. Using the LCFA, we evaluate the absolute probability of a photon producing a pair in each time step. The pair production spectrum for an unpolarized photon per unit time is given by
\cite{PairProductionMeuren,ritus_1985,Berestetskii_b_1989,baier1968quasiclassical}
 \begin{equation}\label{PairProductionSpectrum}
\frac{\text{d}P_{\text{pair}}}{\text{d}y_\gamma \text{d}t}  = \alpha \, \frac{c}{(E_{\gamma}/mc^2)\lambdabar_C}\left\{ \int_{\Tilde{z}}^{\infty} \text{d}t' \text{Ai}(t') \right. \\\left.- \text{Ai}'(\Tilde{z}) \frac{w-2}{\Tilde{z}} \right\},
\end{equation}
where $y_\gamma = E_-/E_\gamma$ is the electron to pair energy ratio, $E_-$ is the produced electron energy, $\Tilde{z} =  [w/\chi_\gamma]^{2/3}$, 
and $\chi_\gamma = \chi \frac{u}{1+u}$ is the quantum non-linearity parameter for the emitted photon. Here the parameter $ w$ is related to the energy ratio $y_\gamma$ as $ w = 1 / (y_\gamma(1-y_\gamma))$. Integrating \cref{PairProductionSpectrum} over $\text{d}y_\gamma$ gives the total pair production probability for an unpolarized photon in a locally constant strong electromagnetic field per unit time, \cite{PairProductionMeuren,ritus_1985,Berestetskii_b_1989,baier1968quasiclassical}
\begin{equation}
    \label{PairProductionProb}
\frac{\text{d}P_{\text{pair}}}{\text{d}t} = -\alpha \, \frac{2c}{3(E_{\gamma}/mc^2)\lambdabar_C}  \int_{4}^{\infty} \text{d}w \, \frac{(2w+1)}{w\sqrt{w(w-4)}} \frac{\text{Ai}'(\Tilde{z})}{\Tilde{z}}.
\end{equation}
When a photon decays, its simulation is terminated, and the propagation of the two created charged particles begins. Their energies are found by randomly picking an energy separation $y_\gamma$ through the pair spectrum in \cref{PairProductionSpectrum}.

The probabilities and spectra (\ref{PhotonProb}--\ref{PairProductionProb}) are often expressed in terms of modified Bessel functions of fractional order, $K_{n/3}$. As examples, see \cite{Matveev1957} and \cite{baier1968quasiclassical,Baie98}. We should note that the definition of the Airy functions used in \cite{ritus_1985} and \cite{Berestetskii_b_1989} differs from that used here by a simple factor ($\pi$ and $\sqrt{\pi}$ respectively). A detailed description of the numerical method for implementation of pair production and photon emission is provided in \cite{NIELSEN2022}, which includes comparisons between sampled spectra and purely theoretical formulas like \cref{PhotonSpectrum}

The trident process, as described in e.g. \cite{FedotovReview2022}, is a two vertex process, which can be  characterized by three terms; a direct term, a two step term, and a cross term. The distinction between the two step and direct terms is the separation between the two vertices (photon emission and pair production). 
Modeling the two step process as two independent processes is a good approximation, this has also been done in other cases (see, for example, \cite{FedotovReview2022,Titov2021}). Consequently, the two step term in the trident process is automatically implemented through the inclusion of separate photon emission followed by pair production in the field. 
In this experiment, we probe values of $\chi$ between $0 < \chi < 2.4$.
Based on investigations, for example \cite{Greger2020,BenKing2018,KingRuhl2013}, the direct term and particularly the cross term will be near negligible in this regime.  
It is possible to determine approximately the relative importance of the direct process and the two-step process by comparing the virtual Weizs\"{a}cker-Williams photon intensity with the real photon intensity. The virtual photon intensity is given by the fine-structure constant up to a logarithmic factor. The real photons have a fairly flat intensity spectrum that scales as $L/X$, where $L$ is the target thickness and $X$ is the effective radiation length, defined as 
\begin{equation}
    \label{eq:effective_rad_length}
    X=\frac{E}{\text{d}E/\text{d}x}
\end{equation}
where $\text{d}E/\text{d}x$ is the energy-loss rate per unit length due to radiation. Therefore, the two processes are comparable in strength for a target thickness of about one percent of the effective radiation length. This is the case for the amorphous setting in our experiment ('random' setting), where $X=X_0=2.30$ cm and $L/X_0 =1.7$ \%. Accordingly, the simulations indicate that $\sim50\%$ of all tridents come from the direct process in the amorphous setting. For the aligned case, the effective radiation length $X$ is significantly shorter than $X_0$ as a result of strong-field effects. Due to stronger radiation, the direct process contributes only a few percent to the total pair rate.
 We choose not to include the cross term and implement the direct term for production in the continuum field through the Weizs\"{a}cker-Williams method of virtual quanta \cite{Jackson_b_1975,Baie98}. This has been investigated in, for example \cite{Greger2020}, and was also used to model the direct process during the early E-144 experiment \cite{SlacE144}. The Weizs\"{a}cker-Williams method deviates significantly from the LCFA method in the high $\chi$ limit \cite{Greger2020,KingRuhl2013} and should deviate approximately $10\%$ from a fully consistent treatment. 
Since the two-step process dominates our experiment, these differences have only marginal influence in our case. However, for future experiments, this difference may provide insight into the importance of both direct and cross terms.

The Weizs\"{a}cker-Williams method integrates a pair-production model with the virtual photon spectrum pertaining to the primary charged particle. In the case of a relativistic particle moving at a constant velocity near the speed of light, the virtual photon spectrum is given by
\cite{Jackson_b_1975}:
 \begin{multline}
   \frac{\text{d}P_{\text{virt}}}{\text{d}x} = \frac{\alpha}{\pi} \left[(K_0\left(\frac{x}{2}\right)K_1\left(\frac{x}{2}\right) \right.\\-\left. \frac{x}{4}\left(K_0^2\left(\frac{x}{2}\right)-K_0^2\left(\frac{x}{2}\right)\right)\right],\label{eq:VirtualPhotonSpectrum}
\end{multline}
where $x = E_\gamma/E$ (we have set $\beta =1$ here and a few other places in \cref{eq:VirtualPhotonSpectrum}) and $K_n$ is a modified Bessel function of the second kind of order $n$. We ignore the influence of the angular variations of the projectile on the spectrum. We further take the direction of the virtual photon to be defined by its instantaneous velocity. It is important to note that our $x$ is different from that appearing in eq. (15.55) in \cite{Jackson_b_1975}: we have applied Jackson's recommended choice for the minimum impact parameter, $b_\mathrm{min} = \hbar/(2mv)$, which is roughly half of the reduced Compton wavelength of the electron, $\lambdabar_C/2$. Using this choice, Jackson's $x$ is half of our $x$ (for $\gamma\gg 1$). It is worth noting that the virtual photon spectrum is extremely sensitive to the choice of $b_\mathrm{min}$ at large values of $x$, while more robust at low values of $x$.
The trident spectrum is found by multiplying the virtual photon spectrum in \cref{eq:VirtualPhotonSpectrum} by the LCFA pair production spectrum \cref{PairProductionSpectrum}:
 \begin{equation}
 \frac{\text{d}P_{\text{WW}}^{\text{trident}}}{\text{d}x\text{d}y\text{d}t} =  \frac{\text{d}P_{\text{virt}}}{\text{d}x} \cdot \frac{\text{d}P_{\text{pair}}}{\text{d}y\text{d}t} =\frac{\text{d}P_{\text{virt}}}{\text{d}x} \cdot \frac{1}{x}\frac{\text{d}P_{\text{pair}}}{\text{d}y_{\gamma}\text{d}t}
  \label{eq:TridentSpectrum}.
\end{equation}
As shown above, $y_\gamma$ has been substituted with the ratio $y= E_- / E$, where $E$ represents the primary particle's energy. This is convenient because $x$ and $y$ are both energies relative to the incoming particle energy. It should be noted that when applying \cref{eq:TridentSpectrum}, we assume locality, that is, both the virtual photon spectrum and the pair-production are evaluated at the projectile's position. In the case of pair production, the argument is that of application of the LCFA. 
In the case of virtual photons, the argument is based on the relatively high photon energies of interest: the reduced Compton wavelength of the electron $\lambdabar_C$ multiplied by the ratio of primary to photon energy is the effective maximum impact parameter.
Even for a 10 GeV photon, where the coherent pair-production probability is much less than the incoherent, this effective maximum impact parameter is still smaller than both the screening radius of the target atoms and the thermal vibration amplitude.
To find the direct trident probability per unit time we integrate \cref{eq:TridentSpectrum} over $\text{d}x$ and $\text{d}y$
 \begin{equation}
  \frac{\text{d}P_{\text{WW}}^{\text{trident}}}{\text{d}t} =  \int^1_0\int^x_0 \frac{\text{d}P_{\text{virt}}}{\text{d}x} \cdot \frac{\text{d}P_{\text{pair}}}{\text{d}y\text{d}t} \text{d}y\text{d}x\label{eq:TridentProbability}.
\end{equation}
This calculation is performed every timestep for every charged particle, so a Chebyshev polynomial \cite{boyd2001chebyshev} is fitted to represent the function as shown in \cite{NIELSEN2022}. Numerical errors introduced by Chebyshev implementations are negligible in comparison to systematic errors introduced in the experiment and misplacement of physical elements during simulation.
When a trident is produced, it is necessary to determine both $x$ and $y$, where we first determine the photon energy $x$, which is then used to determine the pair energy $y$. 
We define the cumulative probability density function and set it equal to a random number $0<r<1$ times the total probability:
\begin{equation}
\label{PropabiblityDensityTridentPhoton}
r\, \frac{\text{d}P_{\text{WW}}^{\text{trident}}}{\text{d}t}
=
 \int^{x}_0\int^{x'}_0 \frac{\text{d}P_{\text{virt}}}{\text{d}x'} \cdot \frac{\text{d}P_{\text{pair}}}{\text{d}y\text{d}t} \text{d}y\text{d}x'.
\end{equation}
By inverting the above expression, we can solve for the ratio $x$. This method allows us to express $x$ as a function of the random number $r$ and $\chi$ that we can fit with a 2-dimensional Chebyshev series $R(t_r,t_{\chi})$.
{As soon as the value of $x$ has been determined, the value of $y$ is calculated in the same manner.
As before, we set the cumulative probability density function equal to a random number multiplied by the total probability for the specific value of $x$ that we have just determined:
\begin{equation}
\label{PropabiblityDensityTridentPair}
r\, \frac{\text{d}P_{\text{WW}}^{\text{trident}}}{\text{d}x\text{d}t}
=
 \int^{y}_0 \frac{\text{d}P_{\text{virt}}}{\text{d}x} \cdot \frac{\text{d}P_{\text{pair}}}{\text{d}y'\text{d}t} \text{d}y'.
\end{equation}

By inverting this function and solving for $y$ as a function of $x$, $\chi$ and $r$, we obtain a three dimensional Chebyshev series. 
The implementation of each Chebyshev series can be found in \cref{sec:AppendixA}.

\subsection{Incoherent processes }\label{sec:AMProcesses}
When a particle penetrates an amorphous material in our setup, electromagnetic processes such as photon emission, pair production, and trident production may occur as a result of incoherent scattering events on atomic nuclei and target electrons. 

These processes depend on the density of the nuclei and electrons inside the amorphous material. The main action of the electrons' is screening of the nuclear charges. They only contribute approximately $1/Z$ times the nuclear contribution, which is an additional 3 \% for germanium. In the case of an aligned crystal, the erratic placement of atomic nuclei as a result of thermal vibrations also results in incoherent scattering contributions to the above-mentioned processes, as does scattering from the target electrons. In an aligned crystal the density of atomic nuclei varies locally according to \cref{eq:nnuclstring}, which is evaluated in each time step for the closest 20 atomic strings inside the crystal. Because the contribution from electrons is small, we simply take it to be proportional to the nuclear contribution.

The photon spectrum from incoherent scattering on target atoms in an amorphous material at high energies, the complete-screening limit, can be calculated using the Bethe-Heitler expression \cite{PDG_2022}
\begin{equation}
    \frac{\text{d}P_{\text{BS}}}{\text{d}t\text{d}E_\gamma} = \frac{1}{X_0E_\gamma}\left(\frac{4}{3}- \frac{4E_\gamma}{3E} + \left(\frac{E_\gamma}{E}\right)^2\right),
    \label{eq:BHrad}
\end{equation}
where $X_0$ is the radiation length; see \cite{PDG_2022} for an expression and values for various materials ($X_0=2.30$ cm for germanium). The radiation length depends inversely on the material density, so when simulating an aligned crystal, we substitute $1/X_0$ by $n_n/(\langle n_n\rangle X_0)$ where $\langle n_n\rangle$ is the average nuclear density of the given material, and $n_n$ is the local nuclear density defined in \cref{eq:nnuclstring}.
It may seem problematic to assume that the radiation is local insofar as screening, and therefore the radius of the target atom, enter the expression for the radiation length and the spectrum (\ref{eq:BHrad}).
 However, the dependence is through a logarithm of the ratio between the effective maximum impact parameter and the minimum as obtained for instance with the Weizs\"{a}cker-Williams approach. The relevant lengths are the screening radius of the target atom and the Compton wavelength of the electron.
Because of their high ratio, more than half of the bremsstrahlung originates from collisions involving impact parameters lower than the amplitude of thermal vibrations.
This justifies the assumption of locality for the distribution (\ref{eq:nnuclstring}). Since the screened nuclear field enters in the determination of both coherent and incoherent spectra, it is possible to have a double counting problem. As discussed elsewhere, the error is the neglect of a modest reduction of the incoherent contribution \cite{Sorensen1996,Nielsen_2020}.

In each timestep, we calculate the probability of emission from
\begin{multline}
        \frac{\text{d}P_{\text{BS-BH}}}{\text{d}t} = \frac{1}{X_0}\left(\frac{4}{3}\ln\left(\frac{E}{E_\mathrm{min}}\right) - \frac{4(E-E_\mathrm{min})}{3E} \right. \\+\left.\frac{(E^2-E_\mathrm{min}^2)}{2E^2}\right),
\end{multline}
where $E_\mathrm{min} = 1$ MeV is a lower energy limit on the photon energy we allow to be emitted. The cutoff is justified because the energy emitted below this region is negligible and any incoming photon with less than 1 MeV has no effect on the experiment. 
When a bremsstrahlung photon is emitted, its energy can be determined using inverse transform sampling where we invert and solve the following expression for $E_\gamma$:
\begin{equation}
\label{eq:BremsstrahlungInversePhoton}
r\, \frac{\text{d}P_{\text{BS-BH}}}{\text{d}t}
=
 \int^{E_\gamma}_{E_\mathrm{min}} \frac{\text{d}P_\text{BS-BH}}{\text{d}t\text{d}E_\gamma'}  \text{d}E_\gamma'.
\end{equation}
We evaluate the inverse as a function of $E$ and fit a Chebyshev series directly to the function.
A description of the Chebyshev implementation can be found in \cref{sec:AppendixA}.

In an amorphous material, high-energy photons produce a pair spectrum given by the Bethe-Heitler formula
\cite{PDG_2022}
\begin{equation}
    \frac{\text{d}P_{\text{PP-BH}}}{\text{d}t\text{d}y_\gamma} = \frac{1}{X_0}\left(1- \frac{4}{3}\left( y_\gamma - y_\gamma^2\right)\right),\label{eq:PairSpectrumInc}
\end{equation}
where $y_\gamma = E_-/E_\gamma$ represents the energy ratio between the produced electron and the decaying photon. We apply this expression also for the incoherent pair-production contribution for aligned or nearly aligned crystals. Remarks similar to those made above for incoherent bremsstrahlung under such conditions apply.
The probability of pair production per unit time is calculated by integrating \cref{eq:PairSpectrumInc} from 0 to 1, which gives the following result:
\begin{equation}
    \label{eq:PairProductionProbabilityInc}
    \frac{\text{d}P_{\text{PP-BH}}}{\text{d}t} = \frac{7c}{9X_0}.
\end{equation}
When a pair is produced by an incoherent process, we are able to determine the pair distribution $y_\gamma$ by inverse transform sampling:
\begin{equation}
\label{eq:PairProductionInverseInc}
r\, \frac{\text{d}P_{\text{PP-BH}}}{\text{d}t}
=
 \int^{y_\gamma}_{0} \frac{\text{d}P_\text{PP-BH}}{\text{d}t\text{d}y_\gamma'}  \text{d}y_\gamma',
\end{equation}
which, in this case, has the following analytical solution:
\begin{multline}
    y_\gamma = \frac{1}{2}+
     \frac{\left(\sqrt{196r^2-196r+81}+14r-7\right)^{1/3}}{2^{4/3}} \\  
     - \frac{2^{1/3}}{\left(\sqrt{196r^2-196r+81}+14r-7\right)^{1/3}}.
\end{multline}

The direct trident spectrum for a relativistic particle colliding with a heavy nucleus, with charge $Z$ in the complete screening limit, has been calculated to lowest order in $Z\alpha$, which is $Z^2\alpha^2$, by Kelner \cite{Kelner1967}. 
Because of the incoherent nature of the problem, the spectrum describes only the direct process and does not have any interference terms with the two-step process as in the coherent case.
The resulting spectrum is the sum of Kelner's equations (26) and (40) in \cite{Kelner1967}, which we have rewritten in terms of $x$ and $y$, by expressing Kelner's parameters as $\xi = y(x-y)/(1-x)$, $\beta = y/x$, and $\zeta = 1-y/x$. The result is
\begin{equation}
    \label{eq:KelnerTrident}
    \frac{\text{d}P_\mathrm{Kel}}{\text{d}x\text{d}y\text{d}t} = \frac{2N_nZ^2\alpha^4}{m^2\pi}\left(\frac{1}{x^2}-\frac{1}{x}\right)(\Phi_A+\Phi_B),
\end{equation}
where $N_n$ is the local atomic density. 
The quantities $\Phi_A$ and $\Phi_B$ are given in \cref{sec:AppendixB}

\begin{figure}[ht]
\centering
		\includegraphics[width=\linewidth]{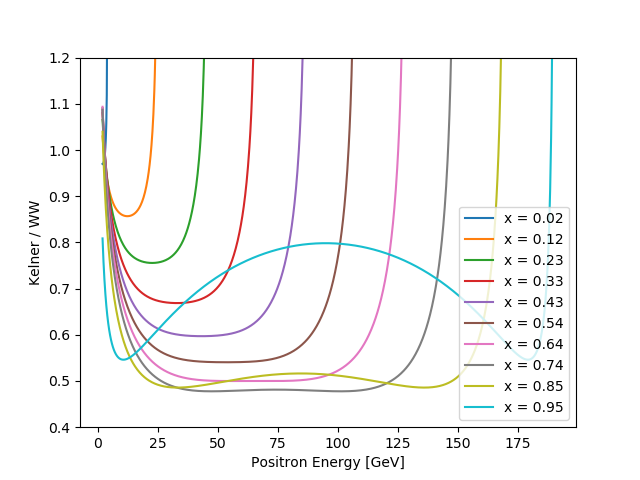} 
\caption{The ratio of the direct trident spectrum computed using the \cref{eq:KelnerTrident} and the direct trident spectrum computed using the WW method of virtual quanta, \cref{eq:TridentSpectrum}, for 200 GeV electrons in amorphous germanium. The incoherent pair-production spectrum, \cref{eq:PairSpectrumInc}, has been substituted for the strong-field pair spectrum. This ratio is shown for different photon energies $x = E_\gamma / E$, where $E = 200$ GeV.}
\label{fig:kelnerVsWW}
\end{figure}

As explained in \cite{Baier2008} and above, the direct contribution to the trident process becomes comparable to the two step contribution when the material is about two orders of magnitude shorter than the radiation length. 
This is the case in our experiment, as in the amorphous case, almost half of all tridents originate from the direct process. 

In \cref{fig:kelnerVsWW}, we show the ratio between Kelner's direct pair spectrum \cref{eq:KelnerTrident} and that determined by the WW method of virtual quanta. The latter is determined by \cref{eq:TridentSpectrum} with the LCFA pair-production spectrum replaced by the incoherent pair spectrum in \cref{eq:PairSpectrumInc}. The ratio is shown as a function of the pair energy ratio $y = E_-/E$, for different photon energies $x = E_\gamma / E$, and it is evident that the two models are not completely consistent. As a result, we use Kelner's method to describe the incoherent direct trident contribution. 
This method is applicable to both amorphous targets as well as to incoherent contributions caused by thermal diffuse scattering in aligned crystals.
 In \cite{Niel_2023}, we noted that the latter incoherent contribution was calculated using the same Weizs\"{a}cker-Williams (WW) method as for the coherent contribution. This is a regrettable error. Also for the simulations made in \cite{Niel_2023}, the incoherent contribution to the direct trident production is determined by Kelner's equations regardless of the target orientation.

 \Cref{fig:kelnerVsWWPhotonIntegrated} shows the ratio of the positron spectra integrated over all photon energies. The various curves represent different choices of the minimum impact parameter used in the WW formula. The value used is $b_\mathrm{min}' = k \cdot b_\mathrm{min}$, where $b_\mathrm{min}$ is the standard choice, effectively $\lambdabar_C/2$, and  the constant $k$ is varied. As shown in \cref{fig:kelnerVsWWPhotonIntegrated}, changing  the minimum impact parameter has an enormous impact on the high-energy tail of the trident spectrum based on the WW approach. With the regular value, $k=1$, the low energy tail agrees within $5\%$ with Kelner's result, but by using $k = 1.25$, there is a much better agreement with Kelner's formula over the main region of interest. As a result, a different choice of the minimum impact parameter, or even an energy dependent version, could be of interest when implementing a simple coherent direct trident model using the WW approach.

\Cref{eq:KelnerTrident} is implemented in a similar manner to \cref{eq:TridentSpectrum}, and the specific Chebyshev implementations are found in \cref{sec:AppendixA}.

\begin{figure}[ht]
\centering
		\includegraphics[width=\linewidth]{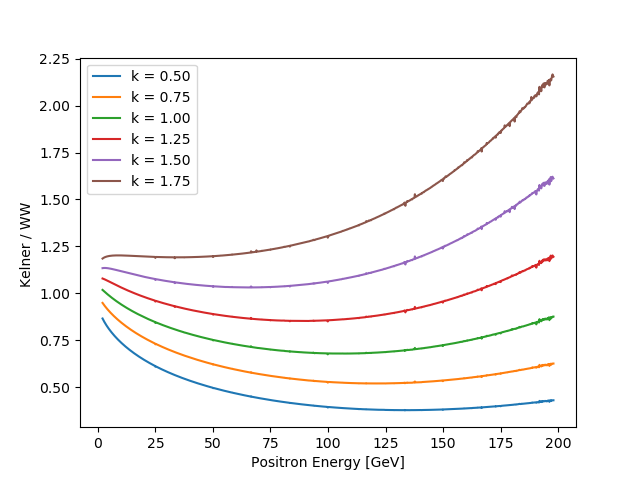} 
\caption{The ratio between the direct pair spectrum integrated over photon energies in \cref{eq:KelnerTrident} and the WW method of virtual quanta in \cref{eq:TridentSpectrum}, where the pair production spectrum in the latter has been exchanged with the incoherent pair spectrum in \cref{eq:PairSpectrumInc}. The various curves represent different factors $k$ multiplied by the $b_\mathrm{min}$ parameter in the WW formula. The initial particle energy is 200 GeV}
\label{fig:kelnerVsWWPhotonIntegrated}
\end{figure}

\section{Data Analysis Algorithms}
The following sections provide details of each step in the data analysis process. The first step is to identify single particle tracks using a tracking algorithm, and then use various matching criteria to identify trident events.

\subsection{Single particle track algorithm}
As mentioned above, the setup is divided into two sections, called  Arm 1 and Arm 2. Specifically, Arm 1 consists of mimosas M1-M5, while Arm 2 consists of M6-M8. Initially, a seed hit in M5 is selected, and then for each hit in M4, a potential track is projected onto M3 by fitting a straight line to the hits in M4 and M5.
We then search for potential hits within a radius of $R = 150$ $\mu$m around the projected hit in M3.
For any hit within the search area, we fit a straight line to the three hits in M3-M5, projecting this line onto M2, searching in an area with radius $R = 350$ $\mu$m repeating the process until a hit in all detectors M1-M5 are used. The search radius in M1 is also $R=350$ $\mu$m.
In the analysis of the deflection angles of low energy tridents, it will be shown that the aligned crystal leads to significant deflection angles for low energy particles. By changing the search radius for M1-M2, a large impact is seen on the low energy part of the spectrum, with the number of accepted tridents increasing by a factor of two when the radius is increased from $R = 150$ to $R= 350$ $\mu$m. The experimental and simulated data shown in \cite{Niel_2023}, are analysed using the conservative $R=150$ $\mu$m for all detectors, whereas $R=350$ $\mu$m has been used in this analysis. For each hit in M5, all permutations of hits in M1-M4 that satisfy the search criteria are investigated, while only the combination of hits that produce the smallest combined distance between the fitted track and the hits used to fit the track is saved. Consequently, each hit in M5 can produce only a single track in Arm 1. By using this algorithm, we avoid the massive number of permutations of hits that can arise when multiple particles travel close together through M3-M5, as in a trident event, for example.

Having produced a list of tracks in Arm 1, we begin working on Arm 2. Arm 2 is designed so that low energy particles most likely will hit M6-M7 and miss M8 because of the large deflection in the magnet. 
M8 was then placed at a greater distance from M7 in order to improve the energy resolution of high energy particles that were deflected very little by M7. 
Therefore, the seed hit in Arm 2 comes from M7, and from this hit we examine all combinations with hits in M6. We project the track onto M8, and if the projection lands within the physical boundary of the detector, we search around the projected hit in a radius $R$ as we did for Arm 1. If the projection is inside the detector and no hit is found, we try another hit in M6, but if the projection lands outside M8, we assume that the hits belong to a low energy track. 

\begin{figure*}[hp]
\centering
\begin{subfigure}{0.73\linewidth}\centering
		\includegraphics[width=\linewidth]{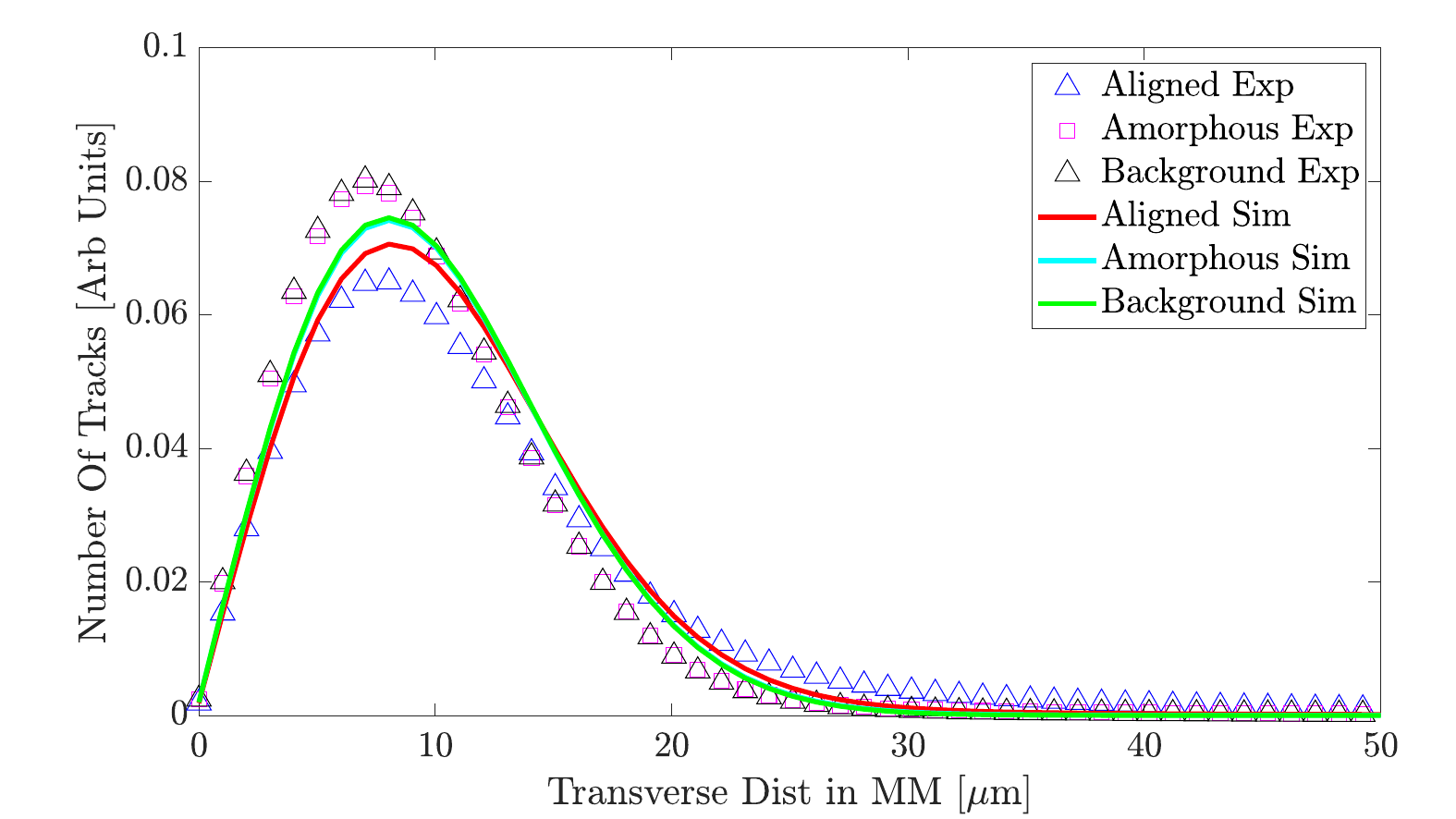} 
\end{subfigure}\\\vspace{35pt}
\begin{subfigure}{1\linewidth}\centering
		\includegraphics[width=\linewidth]{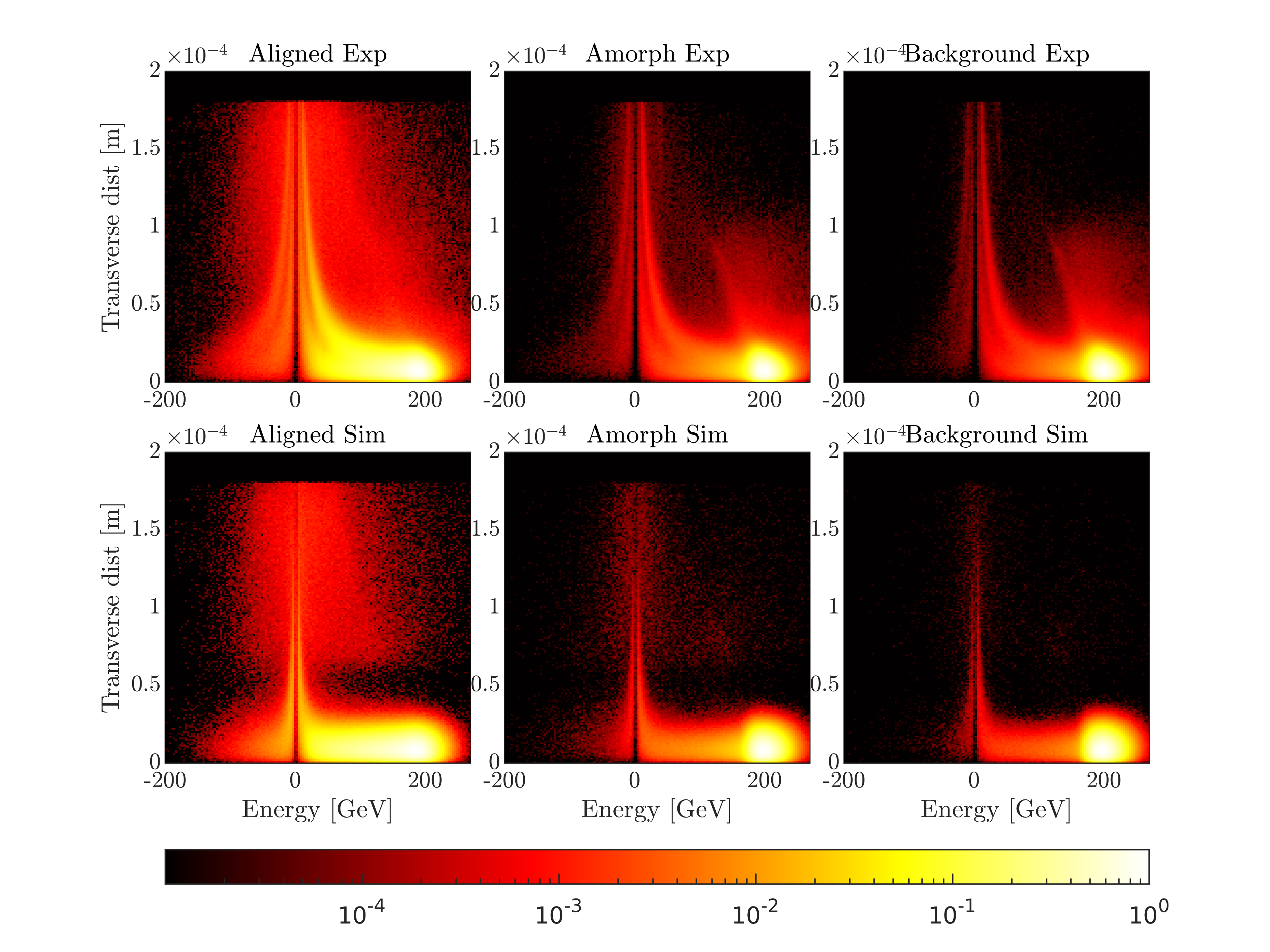} 
\end{subfigure}
\caption{The top figure shows the vertical deflection angle of a single particle track at the center of the magnet. The bottom figure illustrates the relationship between the differential vertical deflection and the track energy. These are the matching criteria to be accepted as a complete single particle track in Arm 1 and Arm 2.}
\label{fig:TransverseDistMMTrackCriteria}
\end{figure*}
\begin{figure*}[hp]
\centering
\begin{subfigure}{0.73\linewidth}\centering
		\includegraphics[width=\linewidth]{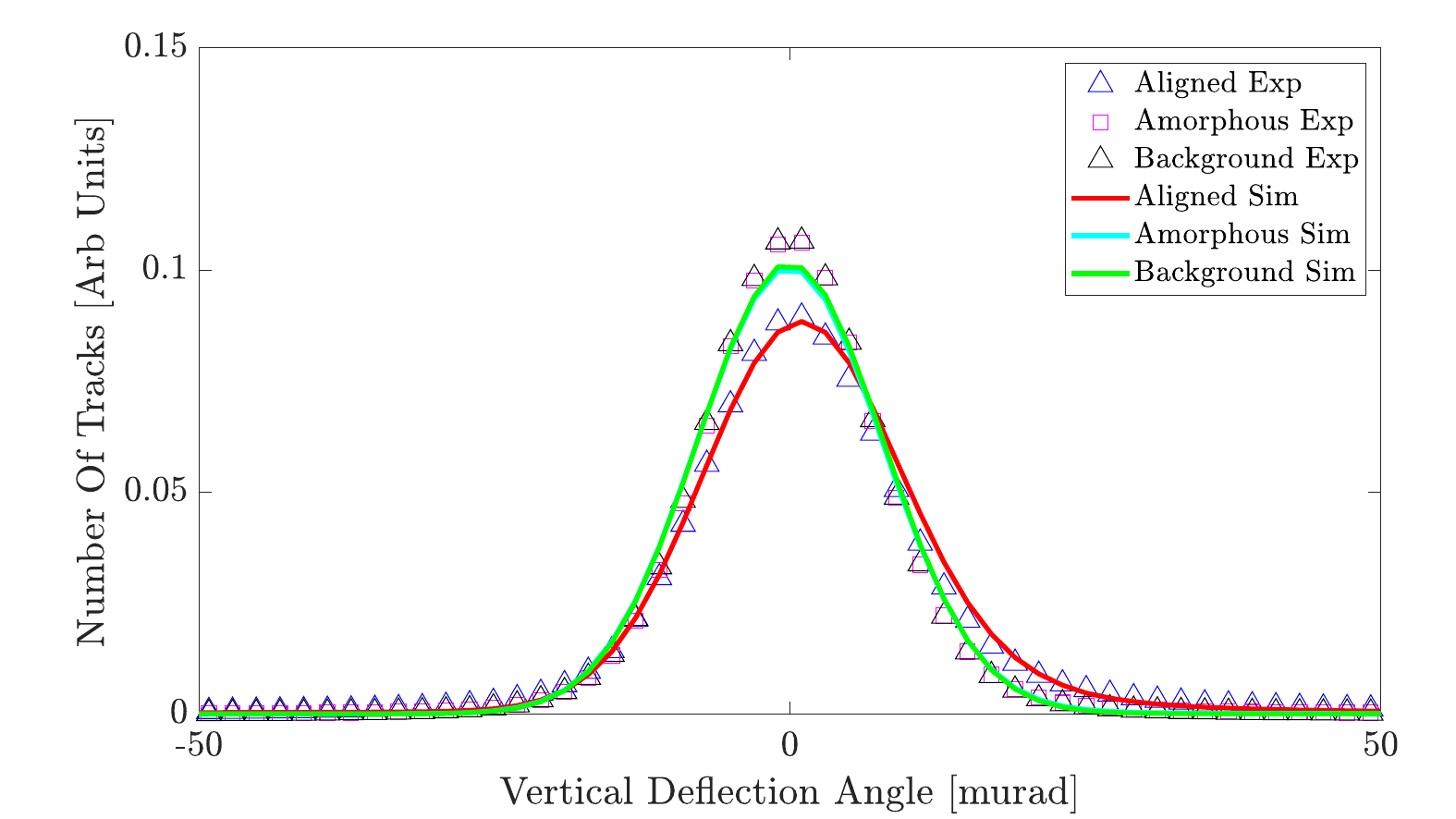} 
\end{subfigure}\\\vspace{37pt}
\begin{subfigure}{1\linewidth}\centering
		\includegraphics[width=\linewidth]{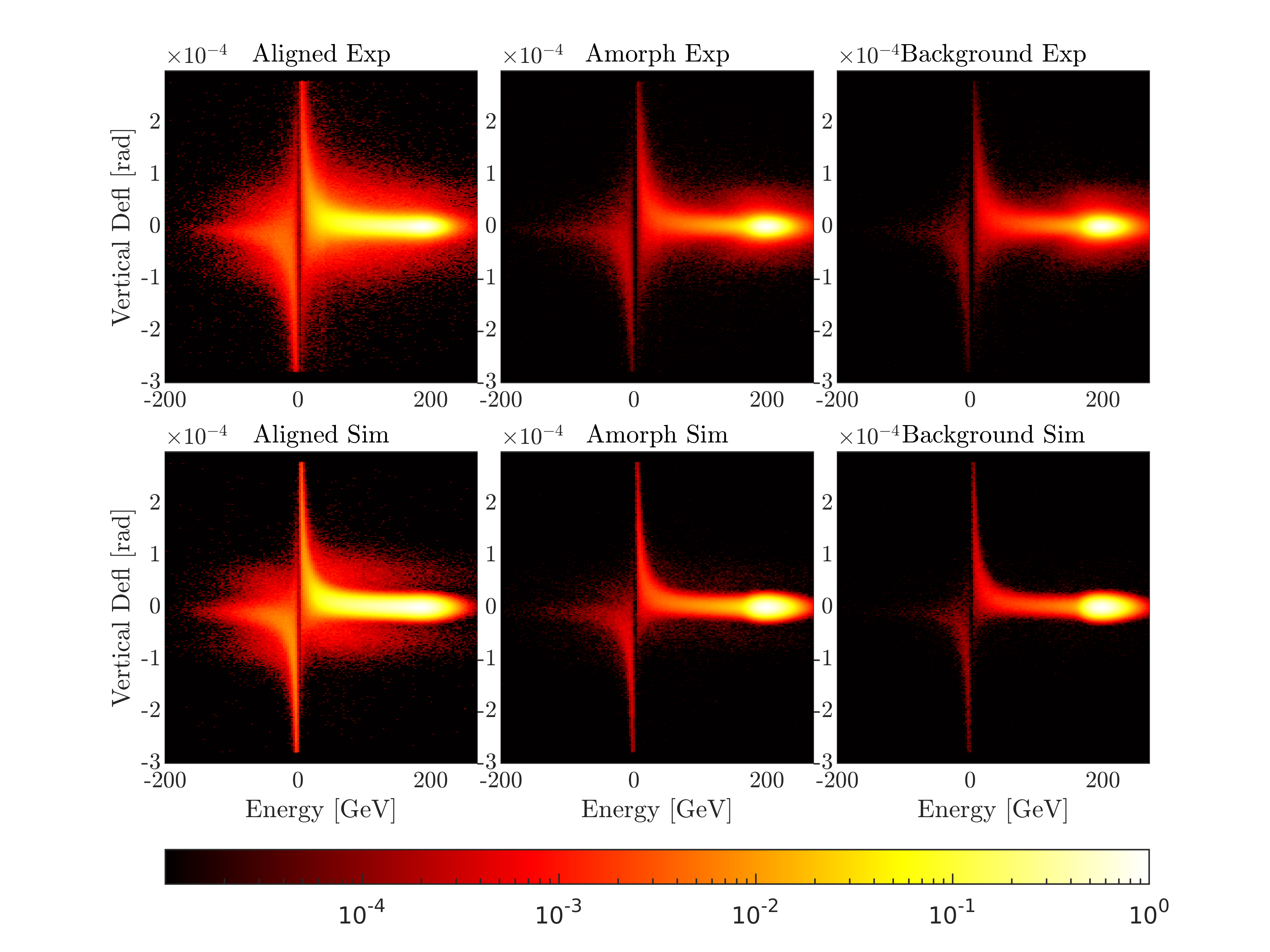} 
\end{subfigure}
\caption{Top figure shows the Transverse distance of a single particle track in the center of the magnet. Bottom figures show the differential Transverse distance vs track energy. These are the matching criteria to be accepted as a complete single particle track in Arm 1 and Arm 2.}
\label{fig:VerticalTrackCriteria}
\end{figure*}
\begin{figure*}[hp]
\centering
\begin{subfigure}{0.73\linewidth}\centering
		\includegraphics[width=\linewidth]{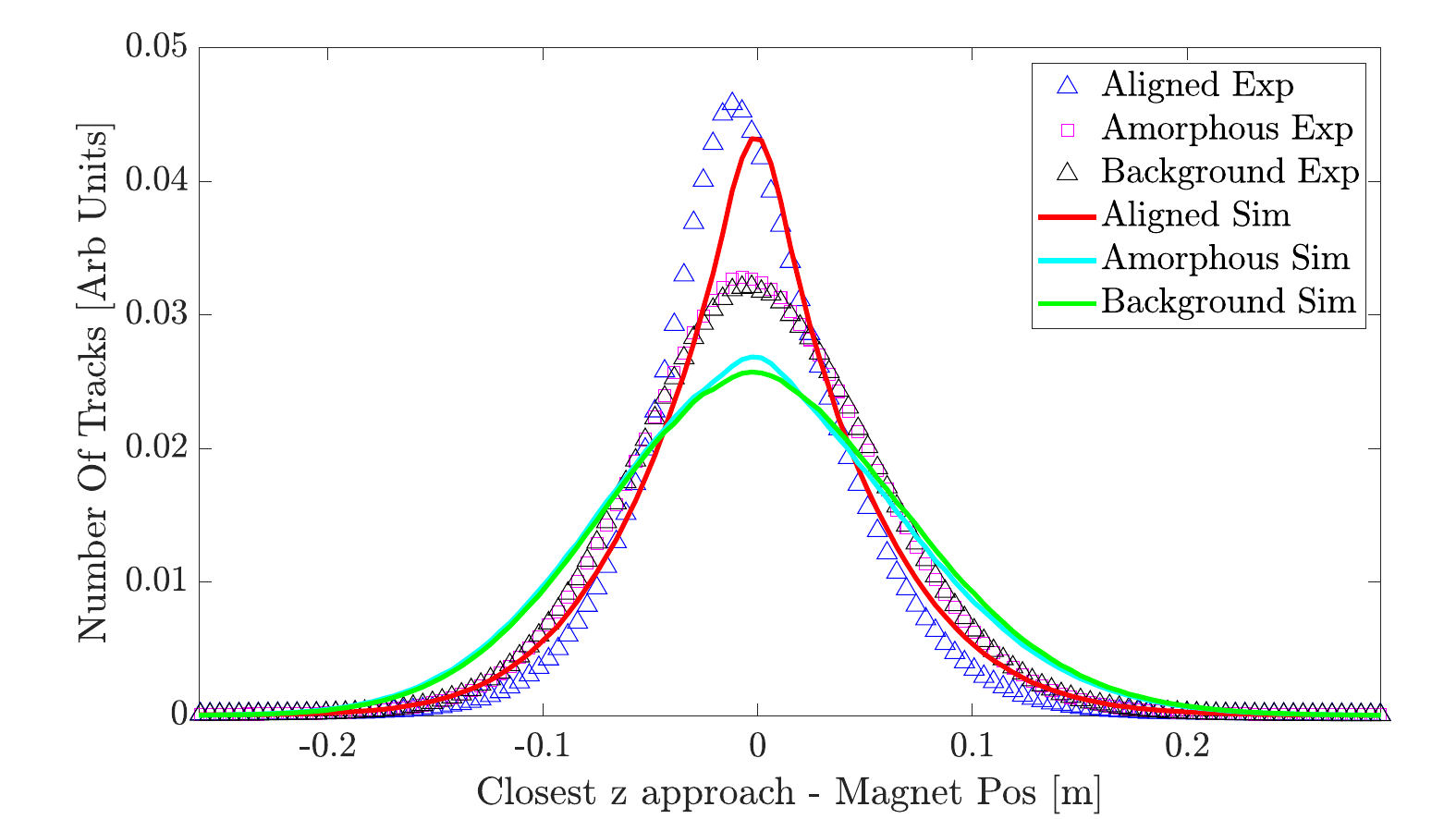} 
\end{subfigure}\\\vspace{35pt}
\begin{subfigure}{1\linewidth}\centering
		\includegraphics[width=\linewidth]{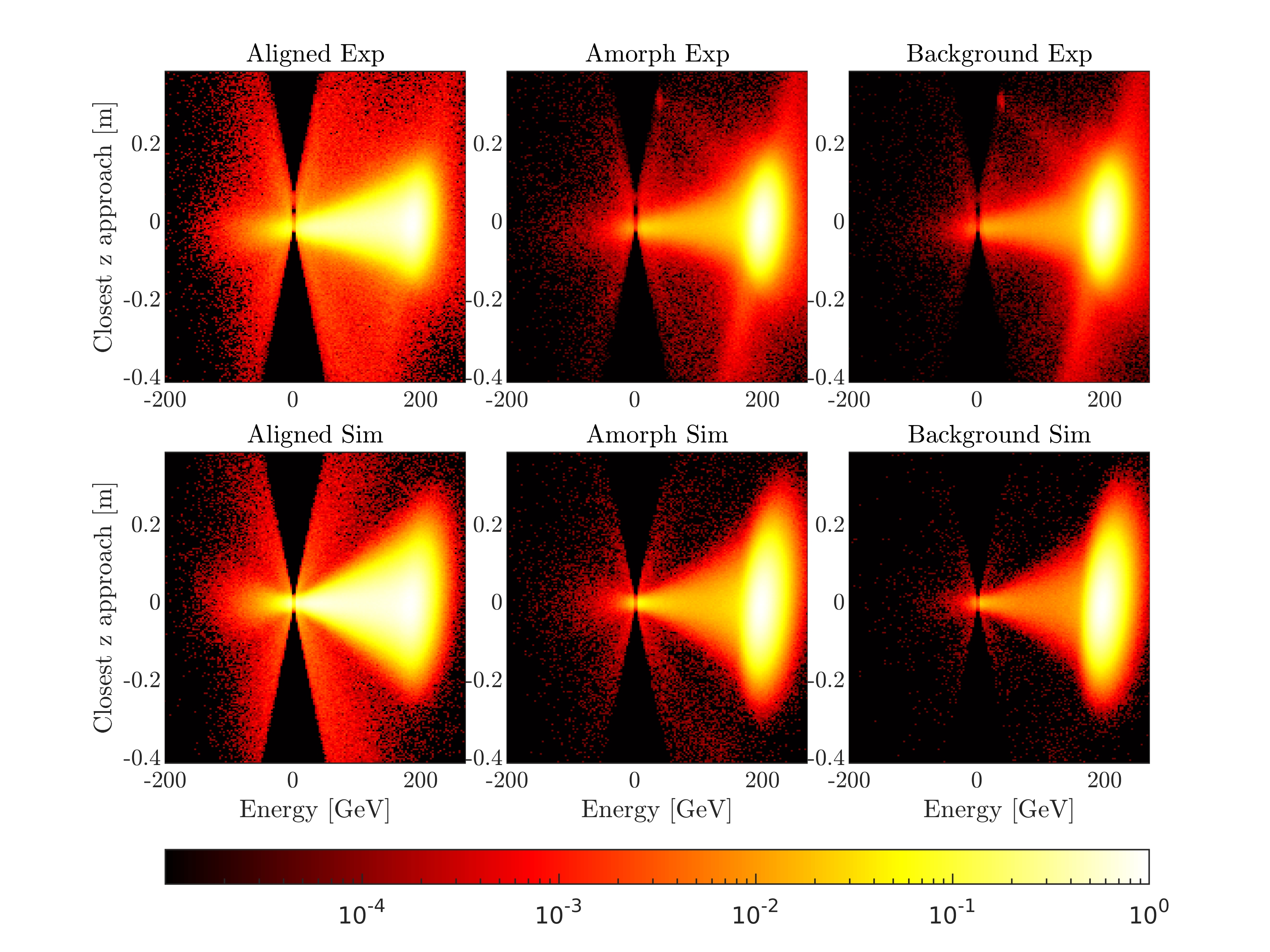} 
\end{subfigure}
\caption{Top figure shows the closest $z$-position of a single particle track between the track in Arm 1 and Arm 2. Bottom figures show the differential closest $z$-position vs track energy. These are the matching criteria to be accepted as a complete single particle track in Arm 1 and Arm 2.}
\label{fig:ClosestZposTrackCriteria}
\end{figure*}

Given the small deflection in the dipole magnet, we can safely assume that the particle deflection occurs instantly in the middle of the Mimosa Magnet (MM). It follows that the particle track should consist of a track in Arms 1 and 2 that crosses in the center of the magnet.
With each hit in M7, tracks in Arm 2 are created and projected into the center of the dipole magnet. The tracks in Arm 2 are matched with tracks in Arm 1 based on the following three criteria:
\begin{itemize}[noitemsep,topsep=5pt,parsep=1pt,partopsep=1pt]
    \item Transverse distance in MM     -     $R_c = 180$ $\mu$m
    \item Vertical deflection angle     -     $\theta_c =  280$ $\mu$rad
    \item $z$ position of closest approach      -     $z_c = 0.5$ m 
\end{itemize}
where distributions of these values for accepted tracks can be seen on \cref{fig:TransverseDistMMTrackCriteria,fig:VerticalTrackCriteria,fig:ClosestZposTrackCriteria} respectively both summed and differential in track energy. All curves are normalized to the total number counts.  The negative energies correspond to positrons and the positive energies correspond to electrons. The numbers indicated in the list above represent the cutoff values for each criterion. We use a track combination that minimizes the transverse distance, which means that we only produce a single complete particle track for each seed hit in M7. 

The transverse distance is the absolute transverse distance between a track in Arm 1 and Arm 2 in the center of the magnet (MM). In general, high energy particles have very low transverse distances, usually less than 20 $\mu$m, as compared to low energy particles which have a greater distance due to scattering. The experimental plots indicate that there is a greater amount of noise, which we believe is the result of the combinatorial nature of the algorithm during the process of building tracks, of which there are more on the electron side (positive energy). It is due to the fact that electrons in trident events are deflected in the same direction, which means if the electrons have similar energies, combinations of hits from both particles might satisfy the matching criteria and produce a complete particle track. We believe this is the causes of the structure in the experimental curves in \cref{fig:ClosestZposTrackCriteria,fig:TransverseDistMMTrackCriteria}, since only electrons display this structure. In the simulated plots,this structure is not visible because the Mimosa software identifies hits based on pixels that are activated when a particle hits. If two particles hit within $50$ $\mu$m on a chip, which is the distance between two pixels with one pixel between them, the simulation combines these two hits into a single hit with their average position. The Mimosa software employs a sophisticated method of deconvoluting hits from clusters of active pixels, which is not implemented in the simulation.

In our case, the vertical deflection should be small, but not exactly zero since the magnet was tilted by $~0.03$ rad in the detector coordinate system. In this case, there was a very small vertical component coming from the magnet, which explains the large cutoff value. The tilt is especially evident in the energy differential plot where the characteristic $1/E$ shape produced by deflection in a magnet with energy $E$ can clearly observed in the low energy tail.  

The closest $z$ approach is the longitudinal position of the closest approach of the track in Arm 1 and Arm 2, with the position of the magnet center subtracted.  Although this value should be zero, it is extremely sensitive to noise for high energy particles due to the small deflection in the dipole, while being less sensitive for low energy particles. Thus, by combining this criteria with the two remaining criteria, which have the opposite sensitivity, we are able to remove non particle tracks from the entire energy spectrum.

 On \cref{fig:ElectronSpectrum}, the energy spectrum of the primary electrons is shown using a logarithmic scale.  
The energy resolution of the magnetic spectrometer for a single particle track was measured to be at $\sigma_E/E \simeq 6.7 \%$ at 200 GeV (including $dp/p \simeq 1\%$ from the beamline). 
The energy resolution is even better for particles with lower energies, due to the larger deflection angle, which is dominated by detector uncertainty until approximately $20$ GeV. 
 In the case of particles below 20 GeV, the uncertainty is dominated by multiple scattering, which is why the setup is surrounded by helium; however, these particles also experience a large deflection, keeping the overall energy resolution below $6.7\%$. 

\subsection{Trident algorithm}
After identifying sets of complete single particle tracks, we now combine the tracks to produce a trident event. 
\begin{itemize}[noitemsep,topsep=5pt,parsep=1pt,partopsep=1pt]
    \item Transverse distance in MM     -     $R_c = 1.5$ mm
    \item Vertical separation angle     -     $\theta_c =  500$ $\mu$rad
    \item Combined energy of trident    -     $E_{cut} = 235$ GeV
    \item One positive charge     
    \item Each track has unique hits in M6-M8
\end{itemize}
It is assumed that only one trident appears in each event; if multiple tridents are identified, we pick the combination of tracks that give the lowest transverse distance in MM. Our findings indicate that minimizing the transverse distance or the vertical separation angle criteria does not make any difference when multiple combinations of tracks can result in a trident event. 
Based on simulated and experimental data for background, amorphous and aligned configurations, \cref{fig:TridentCriteria} illustrates the distributions of the first three criteria for all accepted trident events. The curves are all normalized to the total number of counts.

\begin{figure}[hp!]
\centering
\begin{subfigure}{1\linewidth}\centering
		\includegraphics[width=\linewidth]{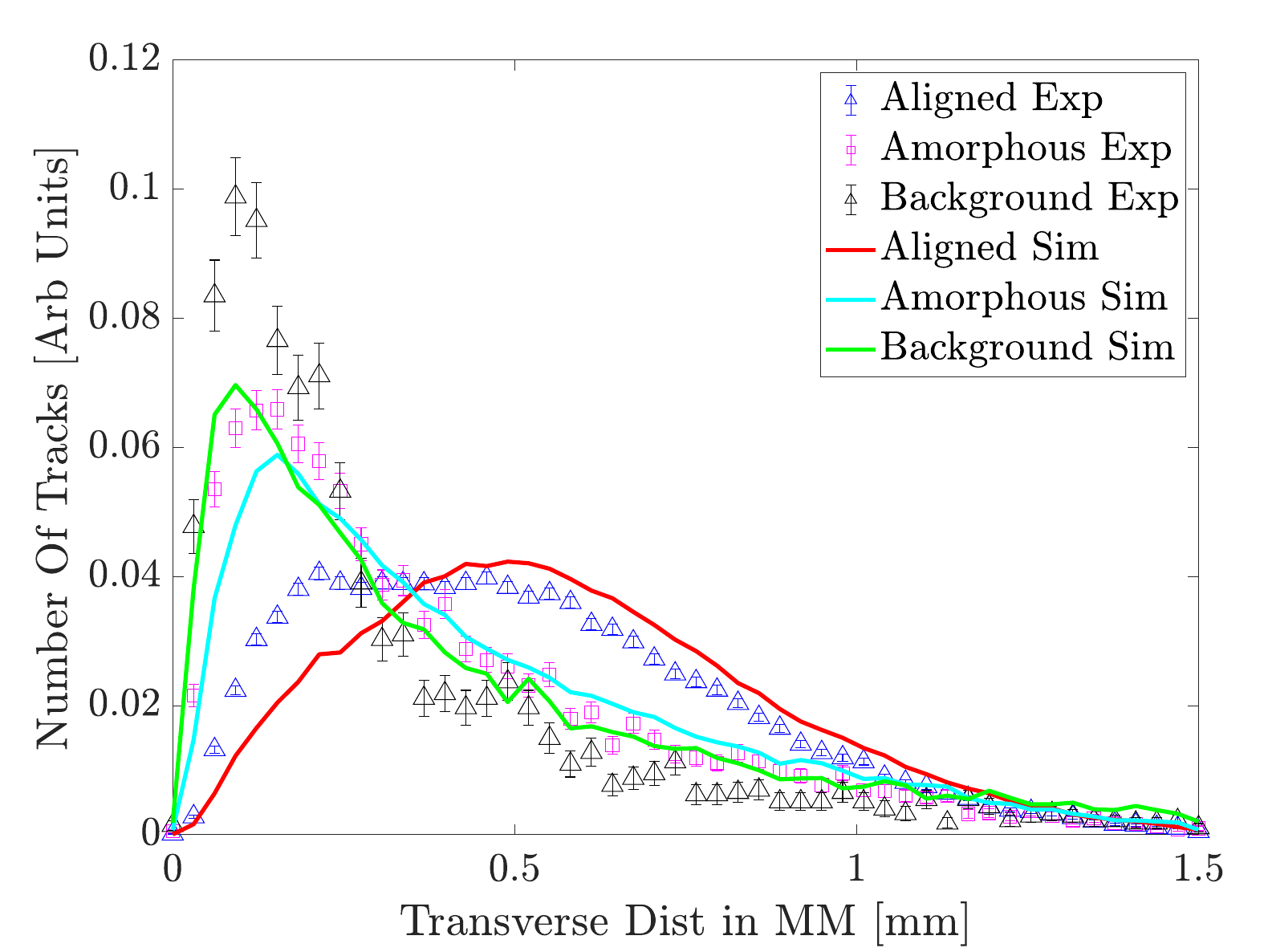} 
\end{subfigure}\\\vspace{10pt}
\begin{subfigure}{1\linewidth}\centering
		\includegraphics[width=\linewidth]{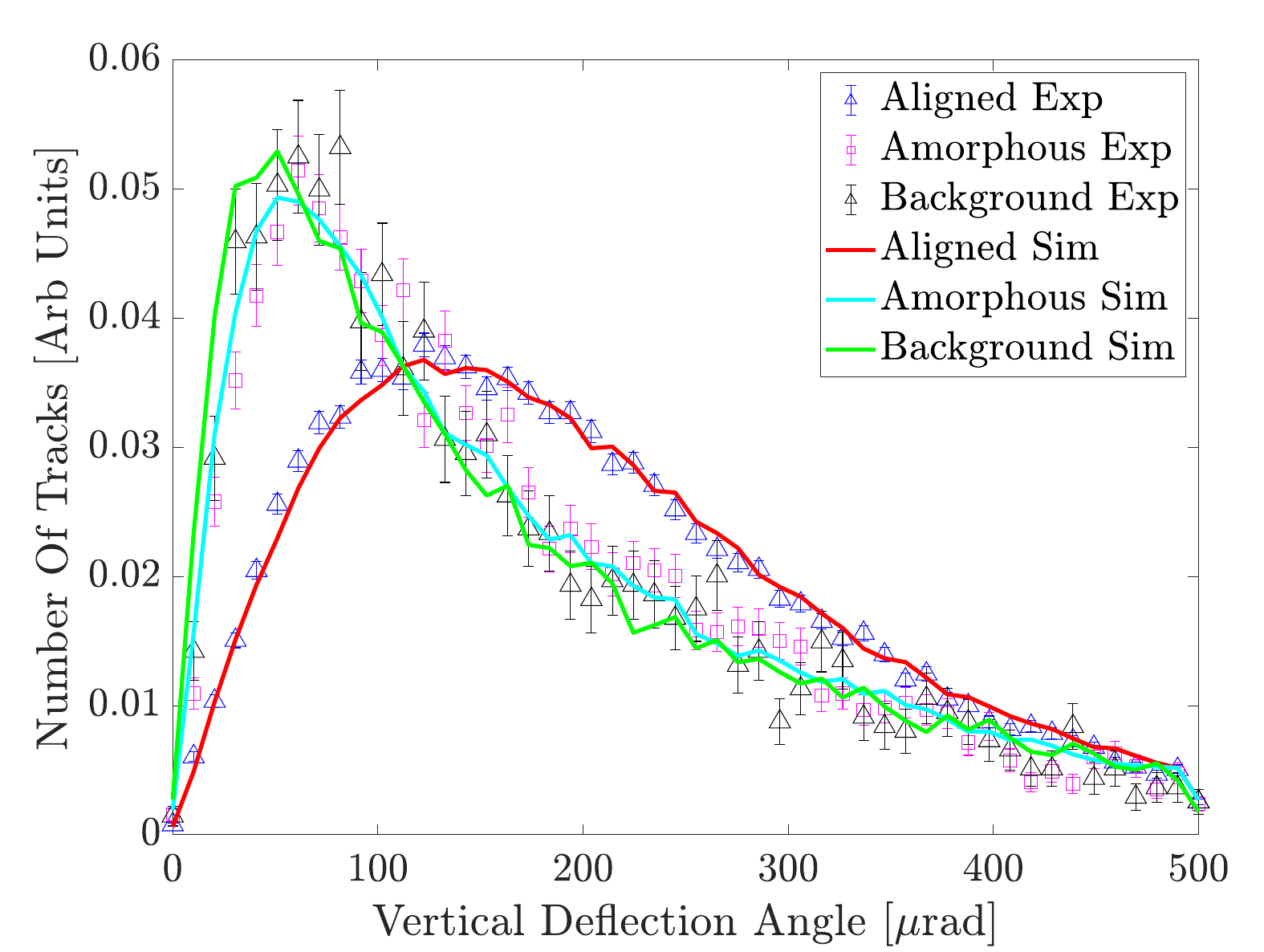} 
\end{subfigure}
\begin{subfigure}{1\linewidth}\centering
		\includegraphics[width=\linewidth]{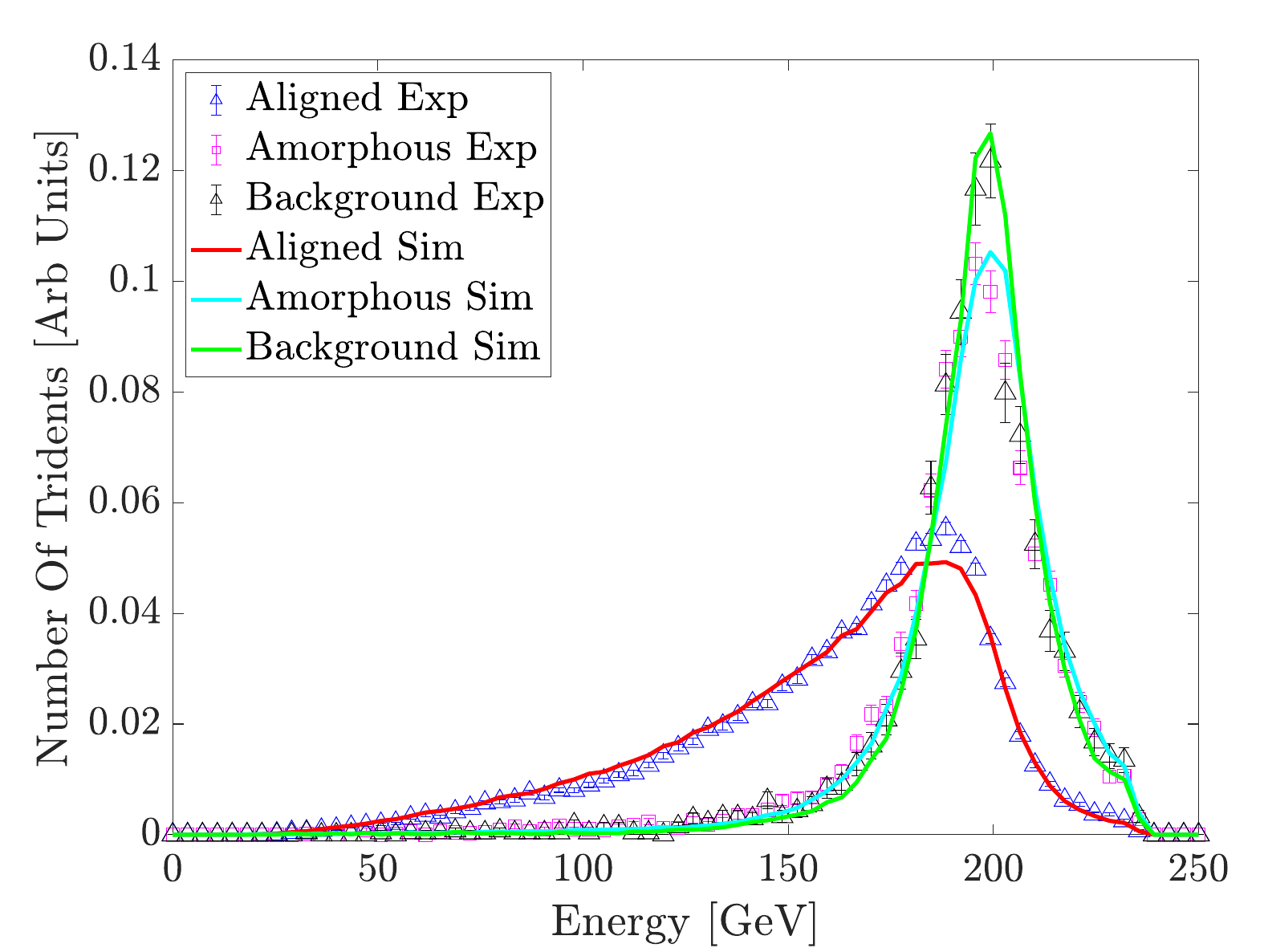} 
\end{subfigure}
\caption{The top figure shows the transverse distance between the three trident tracks from Arm 2 located in the center of the dipole magnet angle for a single particle track. The middle figure shows the vertical separation angle between the three trident tracks from Arm 2. The bottom figure shows the total energy of the three particles that make up a trident. In order to be considered a trident event, three single particle tracks must satisfy these criteria. All curves are normalized to the number of counts.}
\label{fig:TridentCriteria}
\end{figure}

The transverse distance is defined as 
\begin{equation}
    R_\mathrm{trans} = \sqrt{|r_1 - r_2|^2 + |r_1 - r_3|^2 + |r_2 - r_3|^2},
\end{equation}
where $|r_i - r_j|^2$ is the squared absolute distance between the projected hits of Arm 2 track $i$ and $j$ in the  magnet center.
Similarly, the vertical separation angle is defined as 
\begin{equation}
    \theta_\mathrm{vert} = \sqrt{|\theta y_1 - \theta y_2|^2 + |\theta y_1 - \theta y_3|^2 + |\theta y_2 - \theta y_3|^2},
\end{equation}
where $\theta y_i$ is the absolute vertical angle of track $i$ in Arm 2.
  We see that the transverse distance and vertical separation criteria for the Trident algorithm follow the same trend across background, amorphous and aligned configurations. 
The largest values are obtained in the aligned configuration, whereas the second largest values are obtained in amorphous, and the lowest values are obtained in the background configuration. 
 The most notable difference occurs in the aligned configuration since when a pair is formed within an aligned crystal, it occurs in an environment with a very strong electric field. 
 Therefore, when the photon decays, the two particles experience a large force in opposite directions, which results in a transverse momentum of on the order of the critical Lindhard angle \cite{Lind65,JUAnotes}, which in our case is  57 $ \mu$rad for 200 GeV electrons and scales as $1/\sqrt{E}$.
 
 For low energy pairs, which are in abundance, this separation becomes significant, resulting in relatively large values for the transverse distance and vertical separation criteria in the aligned configuration.  In amorphous crystal configurations, the overall material budget in the beam line is larger, resulting in greater scattering of the produced pair than in the background configuration. 
The simulation on average overestimates the transverse distance criteria compared to the experiment, whereas the vertical separation angle is perfectly in agreement. 
 
 The transverse distance is sensitive to the longitudinal position of the magnet center, and we believe that the uncertainty in this measured position is responsible for the slight discrepancy between aligned crystal simulation and experiment. The vertical distance is robust to the magnet's position which is why we see good agreement between experiment and simulation. The total energy criteria ensures that a set of tracks resulting from a combination of hits from the primary and secondary electrons, that would result in a total energy larger than 235 GeV, is discarded. Due to the non zero energy resolution, 200 GeV particles can be measured to have higher energies which is why the cutoff is set at 235 GeV. 
 On \cref{fig:TridentCriteria}, we find remarkable agreement between simulation and experiment for the total energy distribution. The curves appear very similar to the electron spectrum shown on \cref{fig:ElectronSpectrum}. 
 This shows that in the aligned configuration, tridents are often accompanied by photon emission, lowering total energy of the three particles, which is not the case for background and amorphous configuration. The remarkable agreement indicates that subsequent photon emission is accounted for well in the simulation.

\begin{figure*}[ht!]
\centering
\begin{subfigure}{.5\textwidth}
		\includegraphics[width=\linewidth]{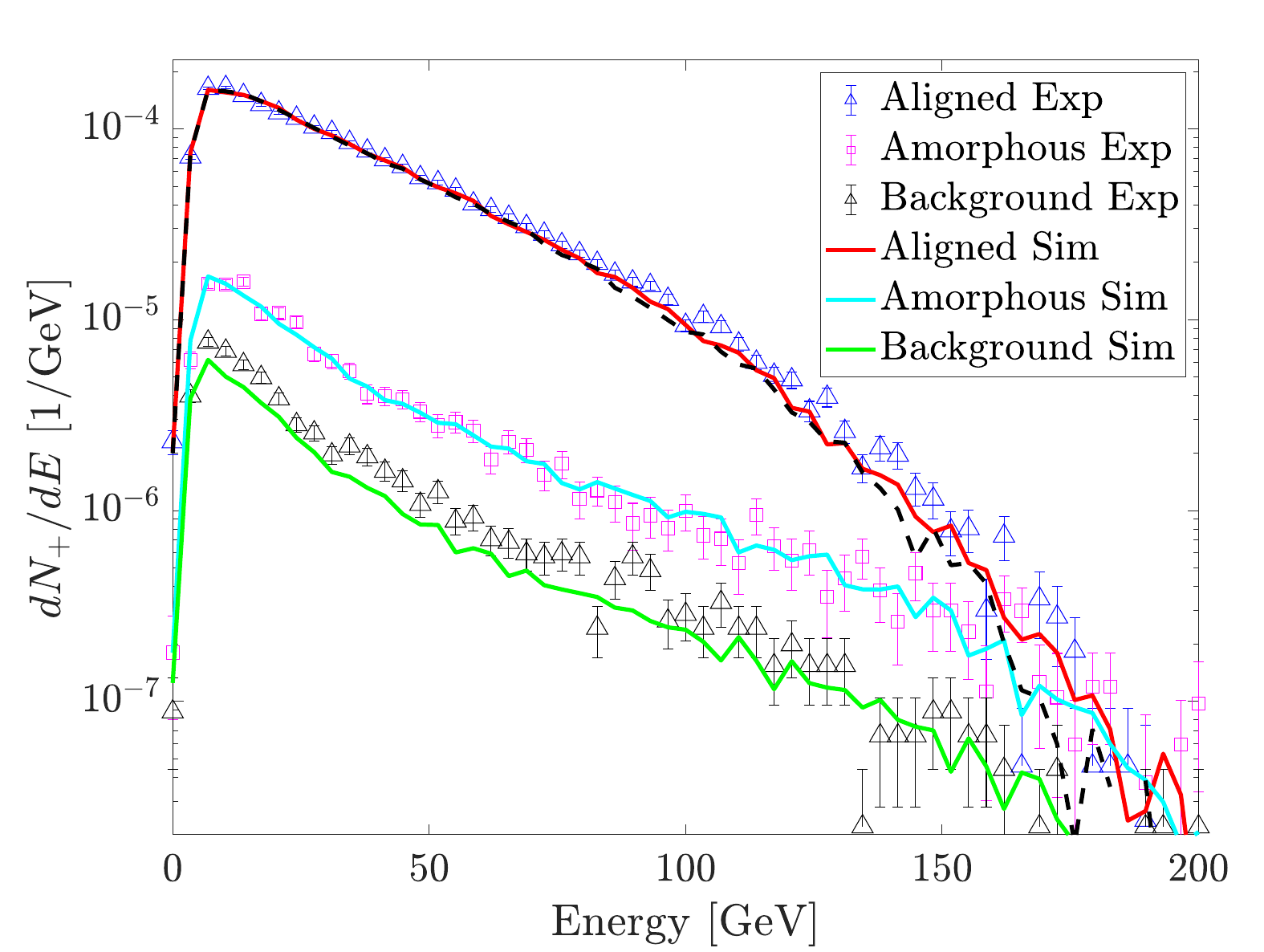} 
\end{subfigure}%
\begin{subfigure}{.5\textwidth}
		\includegraphics[width=\linewidth]{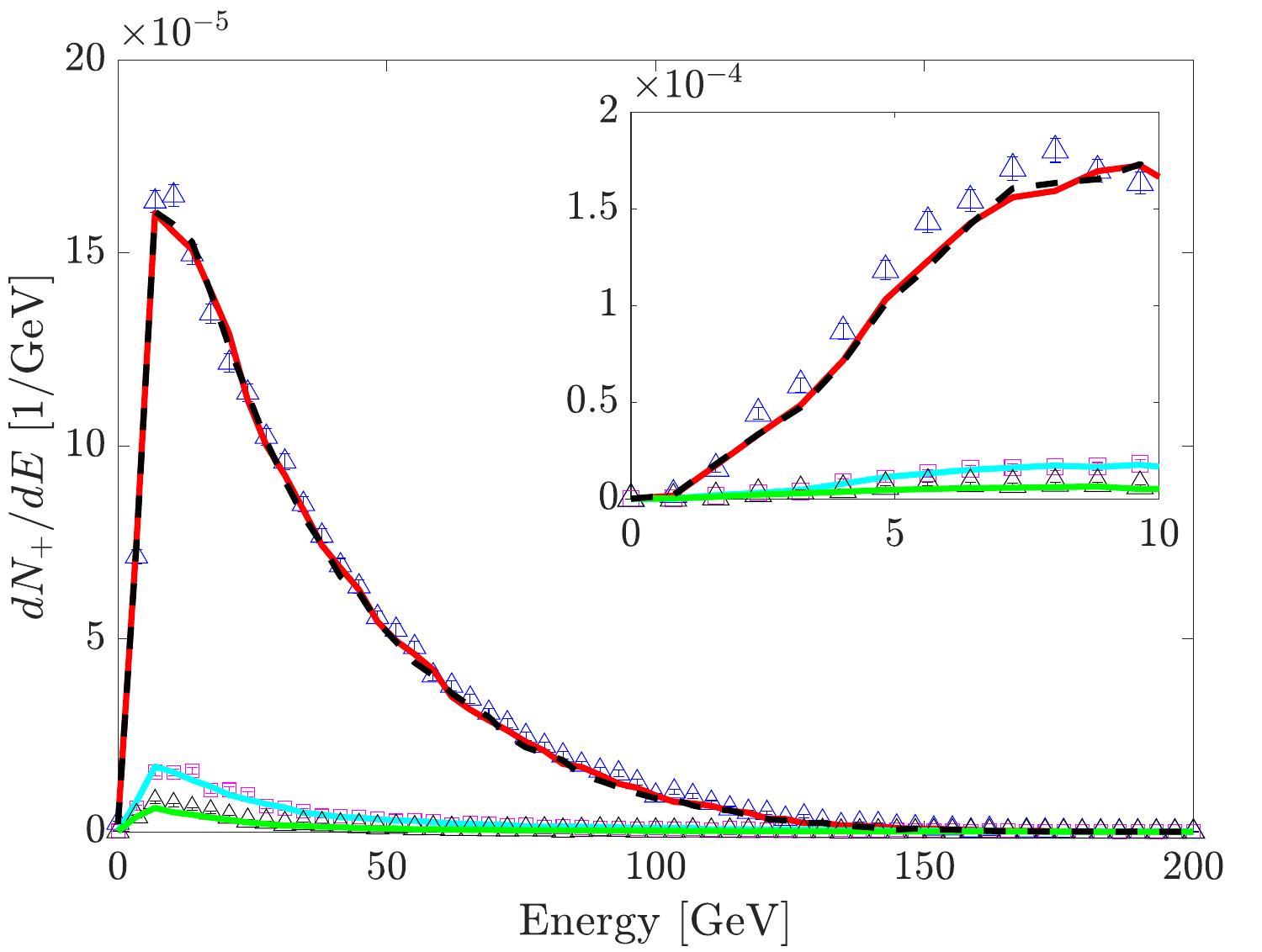} 
\end{subfigure}
\caption{
The positron spectrum of the reconstructed trident events. Solid lines are simulations while squares and triangles are experimental data points. In the aligned orientation, the positron spectrum is blue and red; in the amorphous orientation, the spectrum is magenta and cyan; and in the background, the spectrum is black and green. The figure on the right shows the same data with a linear vertical scale. The black dashed curve represents the simulated positron spectrum in aligned orientation without the direct coherent trident contribution. The inset shows a zoomed-in view of the low energy part, which largely overlaps with the energy range measured in 2007~\cite{Ulrik2009Trident}.}
\label{fig:Spectrum}
\end{figure*}

\section{Theoretical Comparison with Data}
As  mentioned earlier, the theoretical predictions are the result of analyzing a simulated dataset by means of the same data analysis algorithms used to analyze the experimental data. Accordingly, the following and previous comparisons are not the result of fitting, but rather the result of analyzing two independent datasets. Only when comparing absolute rates we need to accurately take into account the efficiency of the setup. This is done by fitting a linear energy-dependent efficiency, $f(E) = aE+b$, to the ratio between the experimental and simulated trident rates for the amorphous case after background subtraction in the region between 17-200 GeV.

When comparing absolute trident rates, the simulated curves obtained from analyzing the simulated datasets are then multiplied by the efficiency factor. The fitting parameters determined by calculating the ratio between the amorphous curves in \cref{fig:Spectrum} are as follows:
$
    a = -0.0006 \pm 0.0016 \text{GeV}^{-1} \quad \text{and} \quad b = 0.98 \pm 0.14$. 
Based on this procedure, the value for $b$ agrees well with expectations, while the value for $a$ is small.  Since we are normalizing to the number of unique single particle tracks, we need to take into account the efficiency of the detectors and expect a value of $b$ of around 1, while the energy dependence is handled by the parameter $a$. Due to the updated alignment procedure mentioned in
\cref{sec:AlignmentOfDetectors},  the efficiency reported in \cite{Niel_2023} differs from the one reported here. In the simulation, the detectors M6-M8 are intentionally misplaced slightly, in the same way as in the experiment, resulting in a slight difference in overall efficiency.An efficiency function is determined by taking a moving average of the direct ratio between each amorphous data point because of the nonlinear energy dependence below 17 GeV.  We can then directly multiply all simulated curves by these nonlinear coefficients below 17 GeV since the bin centers and bin widths are the same for all curves. This is an improvement over what is done in \cite{Niel_2023}, in which the low-energy part of the spectrum is only influenced by the linear energy dependence found by fitting a linear function between 20 and 160 GeV.

A plot of the positron spectrum from trident events is shown in \cref{fig:Spectrum}, where the linear energy efficiency function has been applied to the simulated results. The figure is essentially the same figure as that shown in \cite{Niel_2023}, except for minor changes to the simulated data, together with a change in the acceptance cut for M1-M2 as explained in previous sections. In both the background and aligned configurations, we observe good agreement between the simulated and experimental curves. Compared to the findings in \cite{Niel_2023}, the change in acceptance criteria for M1-M2 results in significant increases in the detection of low energy particles, particularly for the aligned case, where there is a factor 2 increase in accepted tridents. The biggest difference between choosing 150 $\mu$m or 350 $\mu$m in the analysis routine, happens at the single particle track level, where the larger acceptance in M1-M2 allow particles that scatter more heavily in the crystal, to be accepted as a single particle track. The red dashed curve, where the direct coherent trident contribution is omitted, is almost identical to the red full drawn curve, which includes all processes. This should be expected since the crystal thickness is comparable to the effective radiation length (\cref{eq:effective_rad_length}) in the aligned orientation for a 200 GeV electron. On a logarithmic scale, we also observe good agreement across several orders of magnitude. The sharp drop at low energy can be attributed to the fact that low energy particles are deflected outside M6-M7 and cannot be detected due to the setup's detection efficiency. Consequently, we only fit the energy efficiency factor between 17 GeV and 200 GeV.

\begin{figure*}[hp]
\centering
\begin{subfigure}{0.9\linewidth}\centering
		\includegraphics[width=\linewidth]{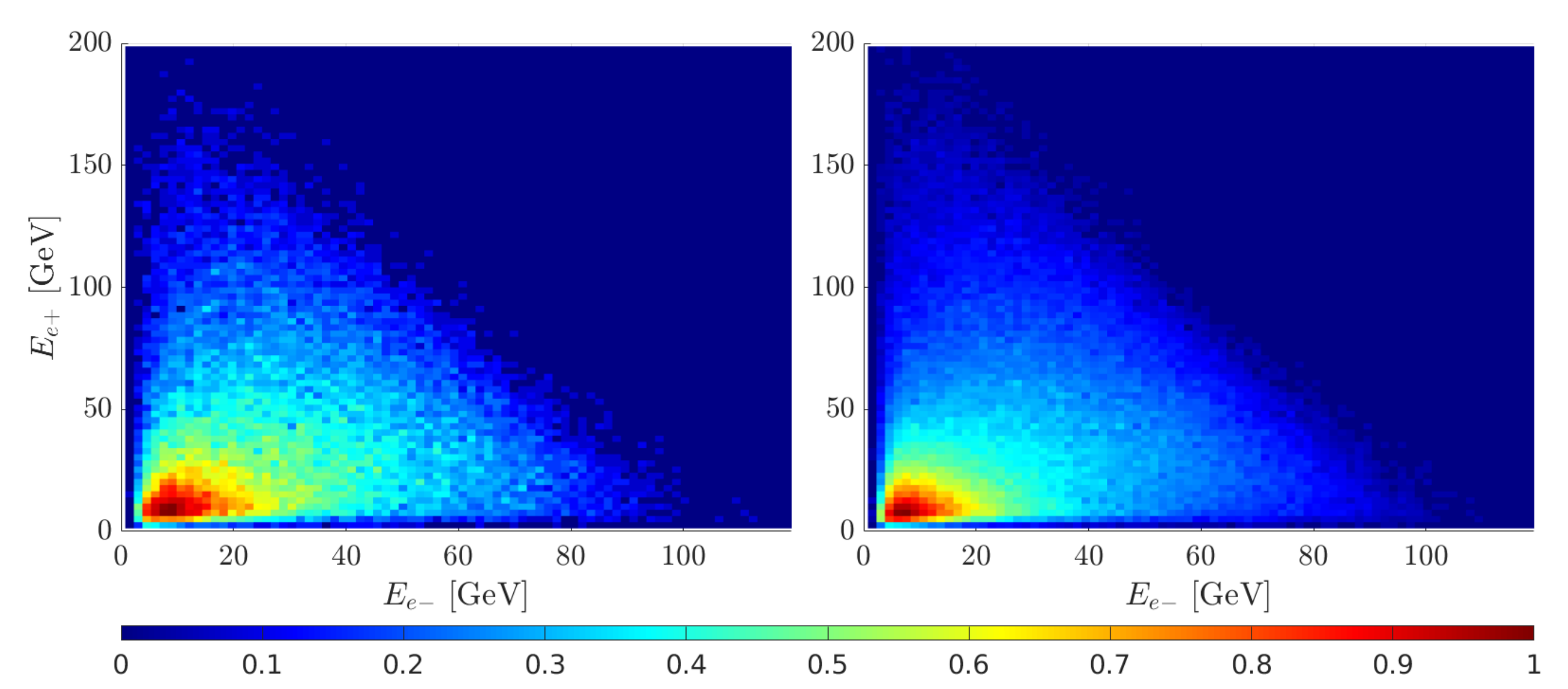} 
  \caption{Trident spectrum differential in positron and electron energy, where the lowest energy electron is used, for the aligned configuration. The figure on the left represents experimental data and the figure on the right represents simulated data. Each plot is normalized to the largest value in the plot and the colors are scaled in relation to the square root of the data point's value.}\label{fig:AlignedDifferential}
\end{subfigure}\\\vspace{10pt}
\begin{subfigure}{1\linewidth}\centering
		\includegraphics[width=0.9\linewidth]{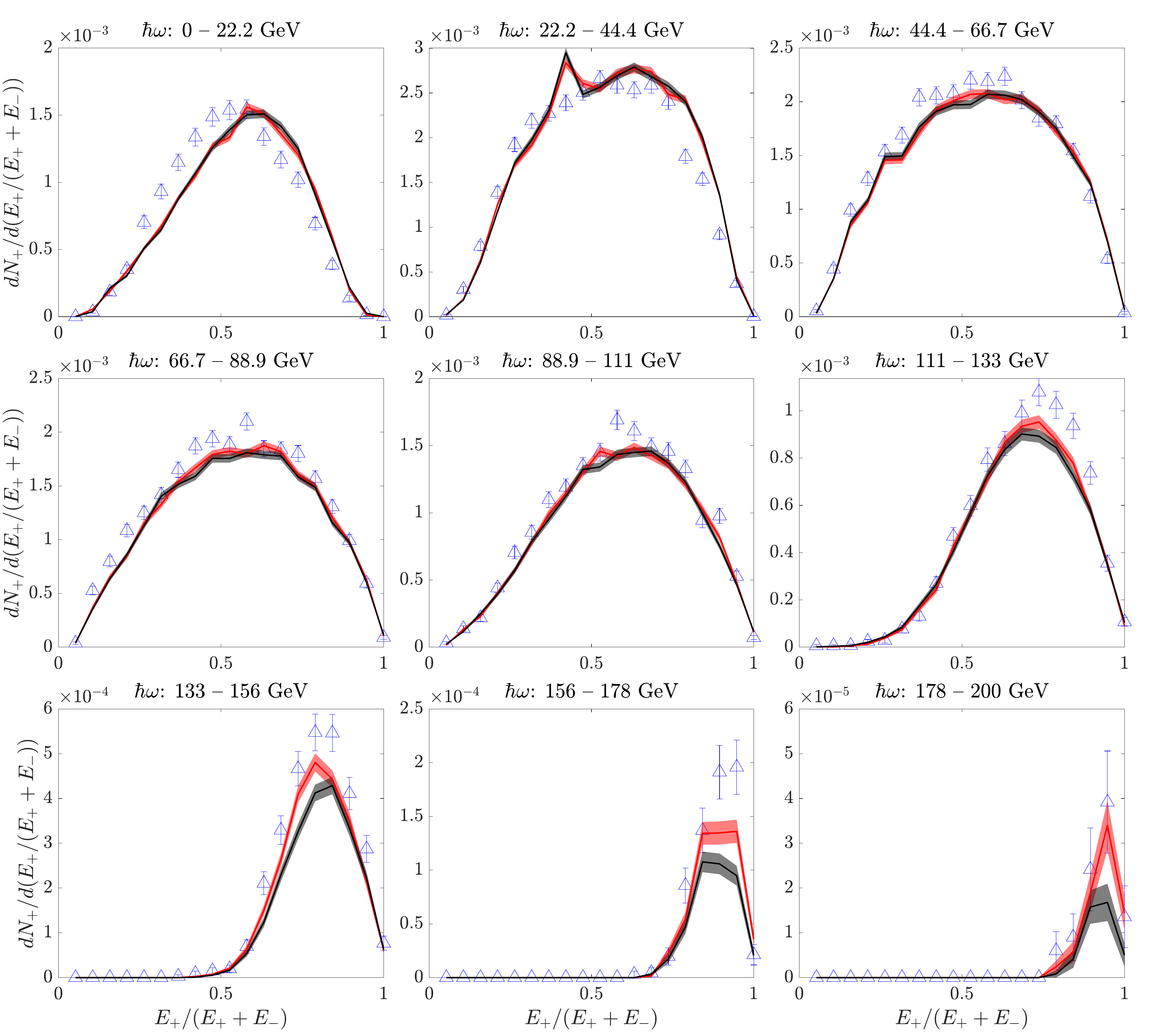}
\caption{The probability of a trident event in the aligned orientation with an energy separation of ($E_{e+}/(E_{e+}+E_{e-})$) between the low energy electron and the positron. The triangles represent experimental data and the solid curves represent simulated data. The red line represents the sum of all trident contributions, while the black line represents no direct coherent trident contribution. In each frame, a specific photon energy is represented (sum of electron and positron energy).  Colored areas indicate a statistical error band of one $\sigma$ around the simulated curves.}
		\label{fig:AlignedSeparation}
\end{subfigure}
\caption{\phantom{asdfasdf}}
\end{figure*}
\begin{figure*}[hp]
\centering
\begin{subfigure}{0.9\linewidth}\centering
		\includegraphics[width=\linewidth]{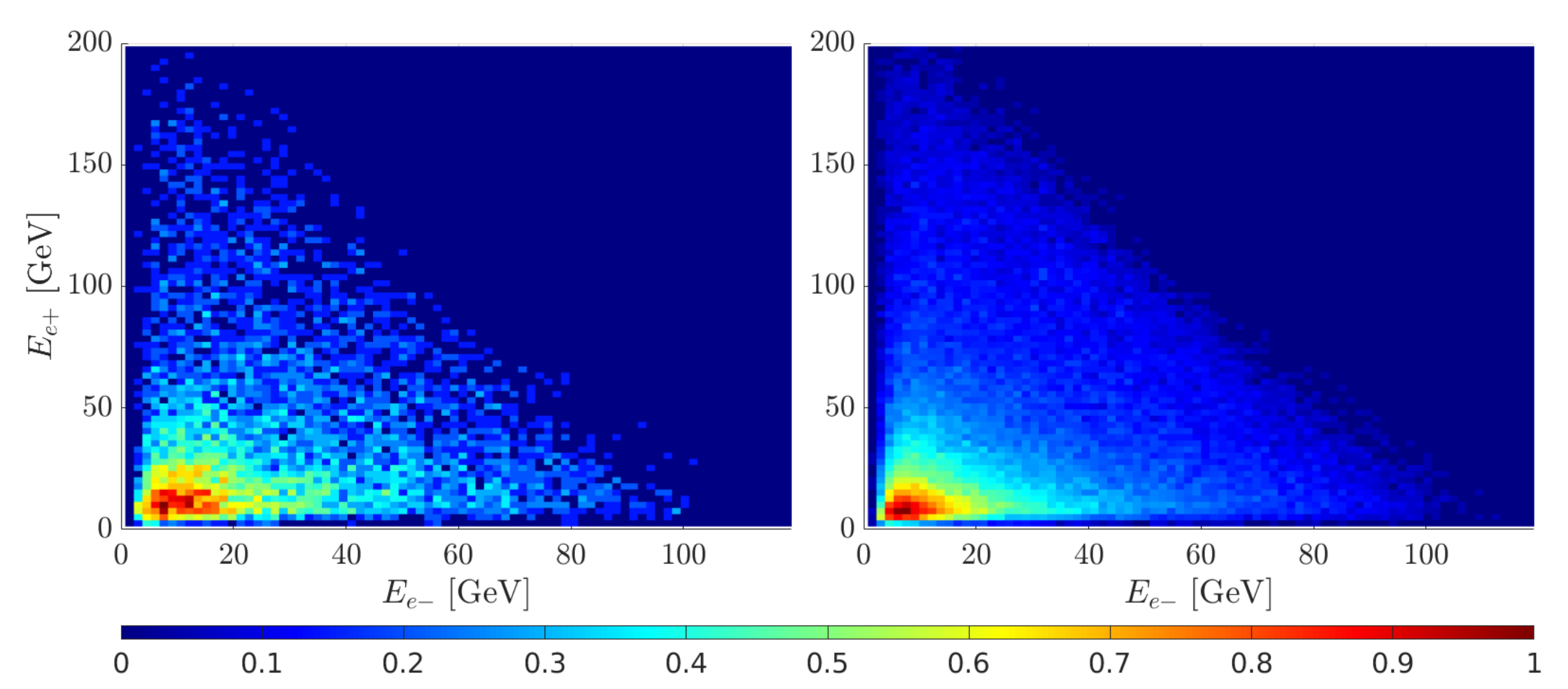} 
\caption{Same as  \cref{fig:AlignedDifferential}, but for amorphous orientation. }
\label{fig:AmorphDifferential}
\end{subfigure}\\\vspace{10pt}
\begin{subfigure}{1\linewidth}\centering
		\includegraphics[width=0.9\linewidth]{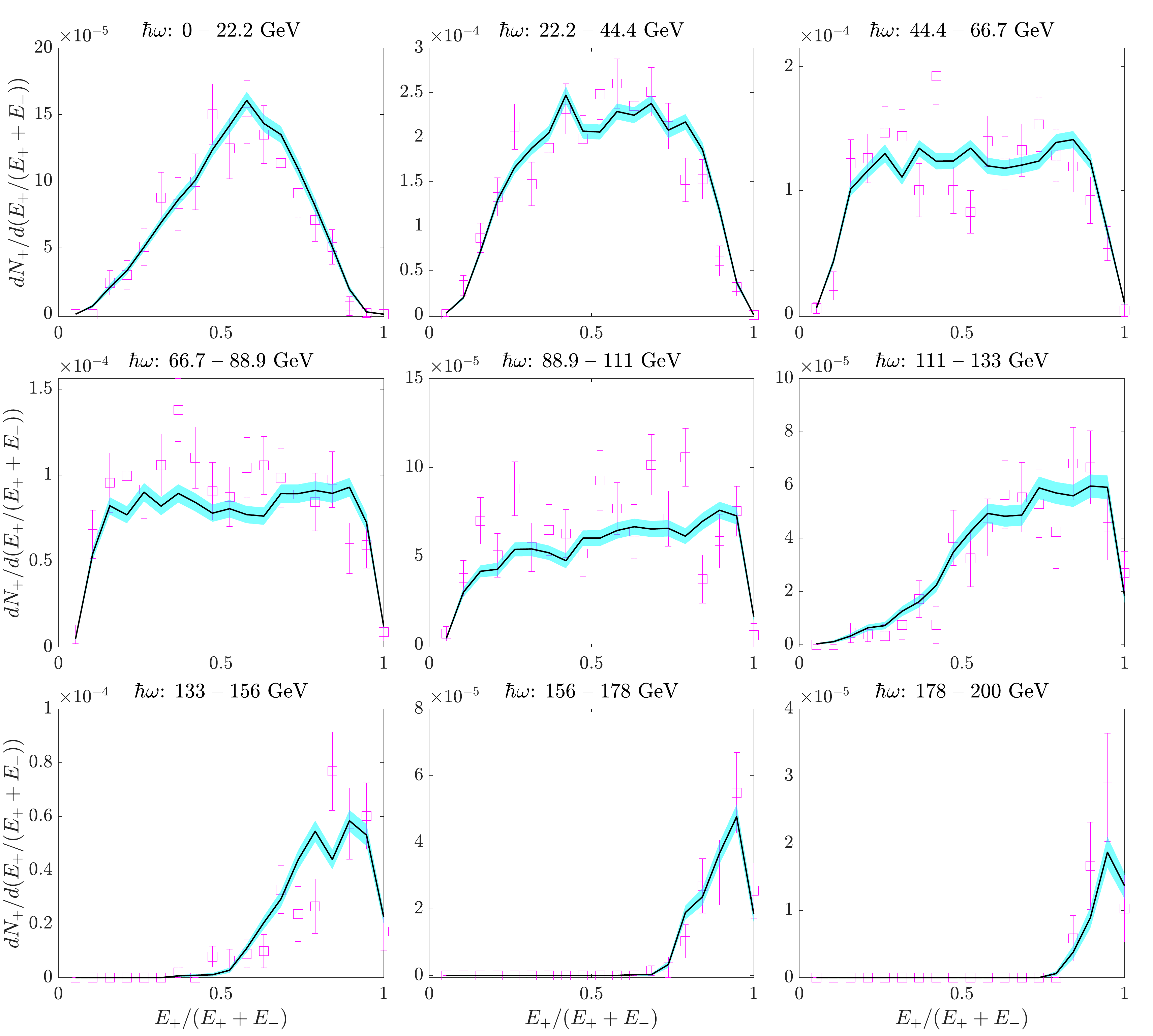}
\caption{Same as  \cref{fig:AlignedSeparation}, but for amorphous orientation. }
		\label{fig:AmorphSeparation}
\end{subfigure}
\end{figure*}
\begin{figure*}[hp]
\centering
\begin{subfigure}{0.9\linewidth}\centering
		\includegraphics[width=\linewidth]{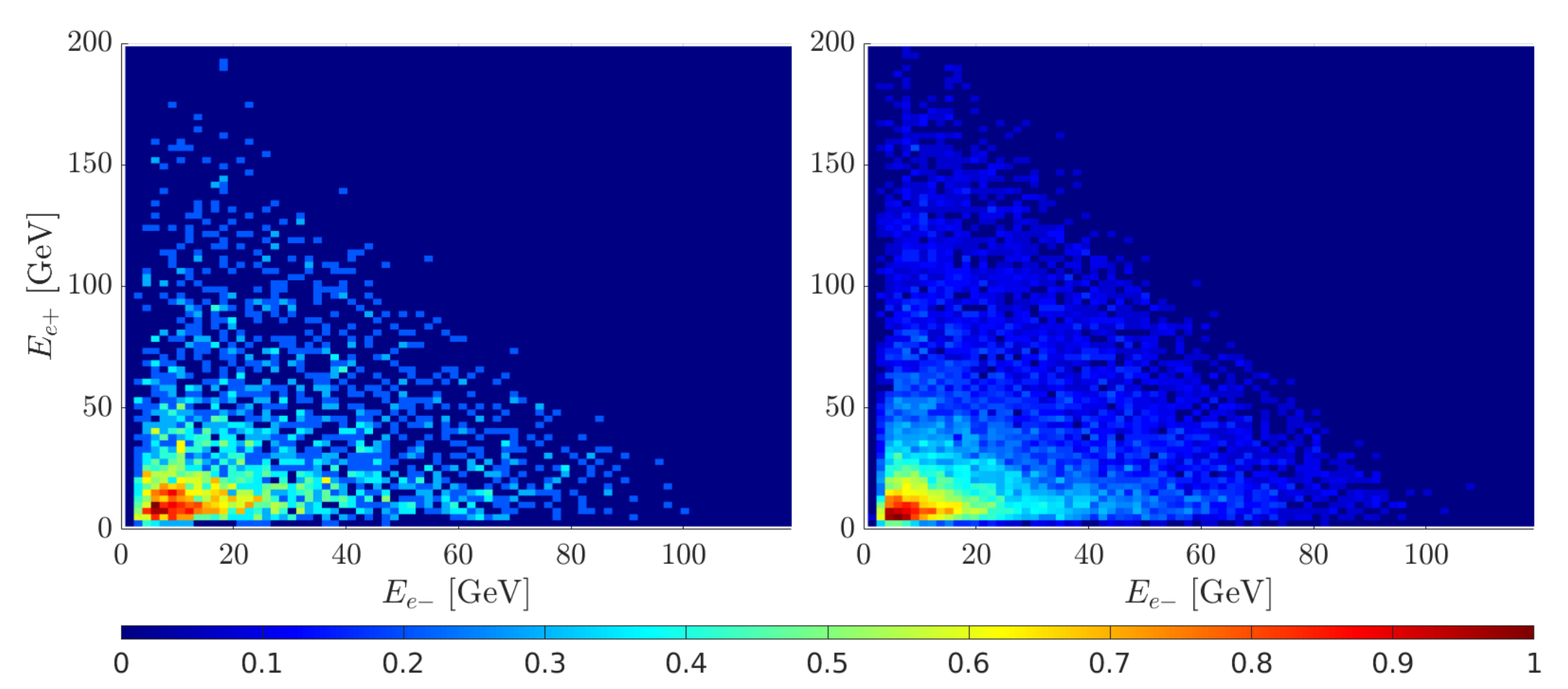} 
\caption{Same as  \cref{fig:AlignedDifferential}, but for background with no crystal. }
\label{fig:BackgroundDifferential}
\end{subfigure}\\\vspace{10pt}
\begin{subfigure}{1\linewidth}\centering
		\includegraphics[width=0.9\linewidth]{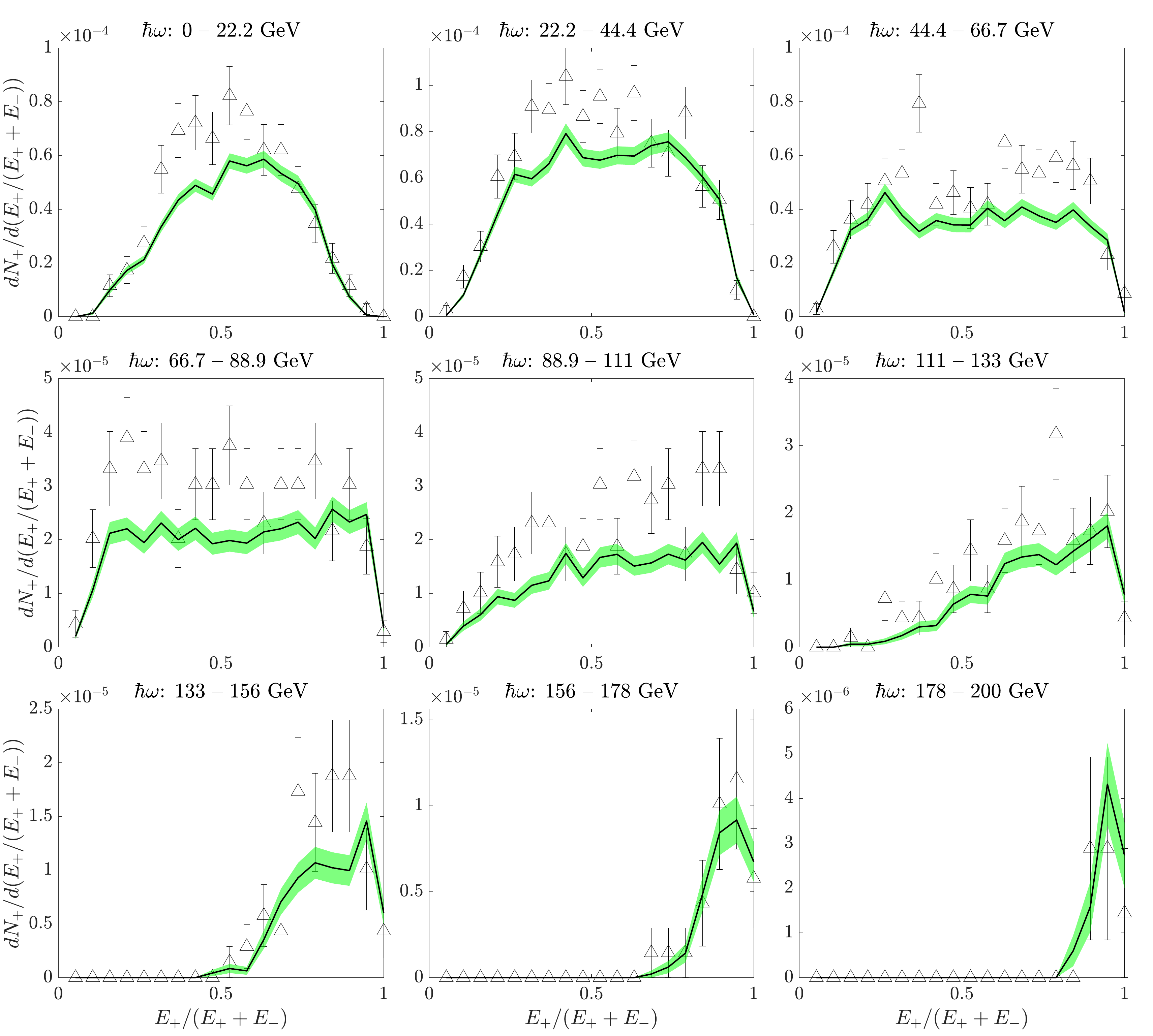}
\caption{Same as  \cref{fig:AlignedSeparation}, but for background with no crystal. }
		\label{fig:BackgroundSeparation}
\end{subfigure}
\end{figure*}

The \cref{fig:AlignedDifferential,fig:AmorphDifferential,fig:BackgroundDifferential} show trident spectra for aligned, amorphous and background configurations respectively, differential in positron and electron energy, using the lowest energy electron.   
The lowest energy electron rarely exceeds 100 GeV, which only happens due to the energy uncertainty, and the axis is cut off at this point. The structure across experimental and simulated surfaces agree well and we see the symmetric energy distribution between the positron and electron for low pair energies. The energy distribution is tilted as the pair energy increases because the lowest energy electron in the trident event is used for these plots. The fact that we use the lowest energy electron in combination with the total trident energy cut of 235 GeV, results in the sharp cutoff that goes from 200 GeV on the positron axis to 100 GeV on the electron axis. Due to the finite uncertainty of a particle hit in the Mimosas, we have an energy uncertainty which can lead to energies larger than 200 GeV for a single particle. We see that the amorphous and background tridents are more localized at low pair energies compared to the aligned case, which extends to higher pair energies. This is expected as the background and amorphous configurations only contain incoherent processes, whereas the aligned configuration is dominated by coherent contributions, that extends further into the pair-energy spectrum. This is clearly visible on \cref{fig:TridentTypesPhotonEnergy}, where all tridents that are produced in the simulation are included, and not just the ones that are detected.

\begin{figure}[hp!]
\centering
\begin{subfigure}{1\linewidth}\centering
		\includegraphics[width=\linewidth]{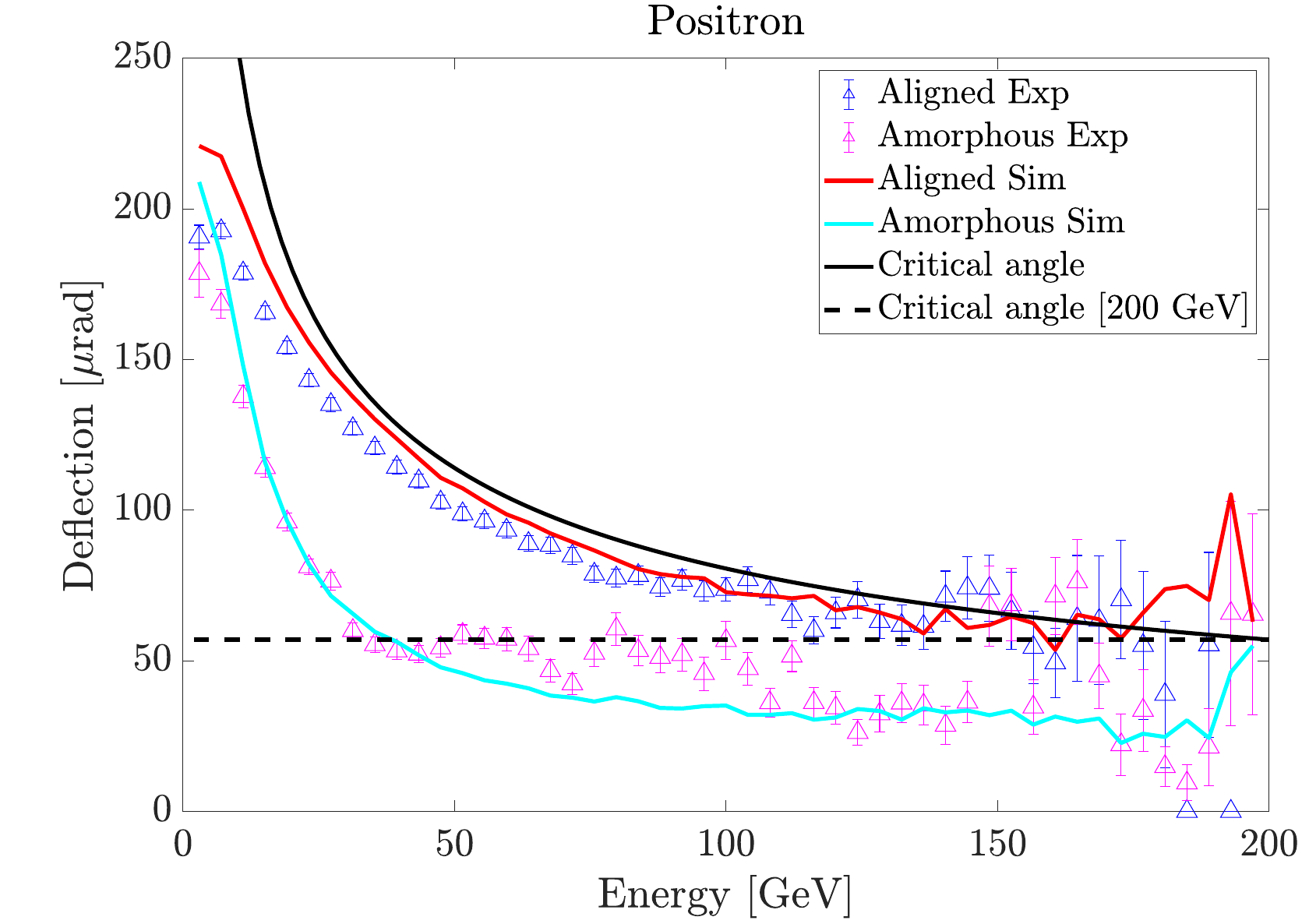} 
\end{subfigure}\\\vspace{30pt}
\begin{subfigure}{1\linewidth}\centering
		\includegraphics[width=\linewidth]{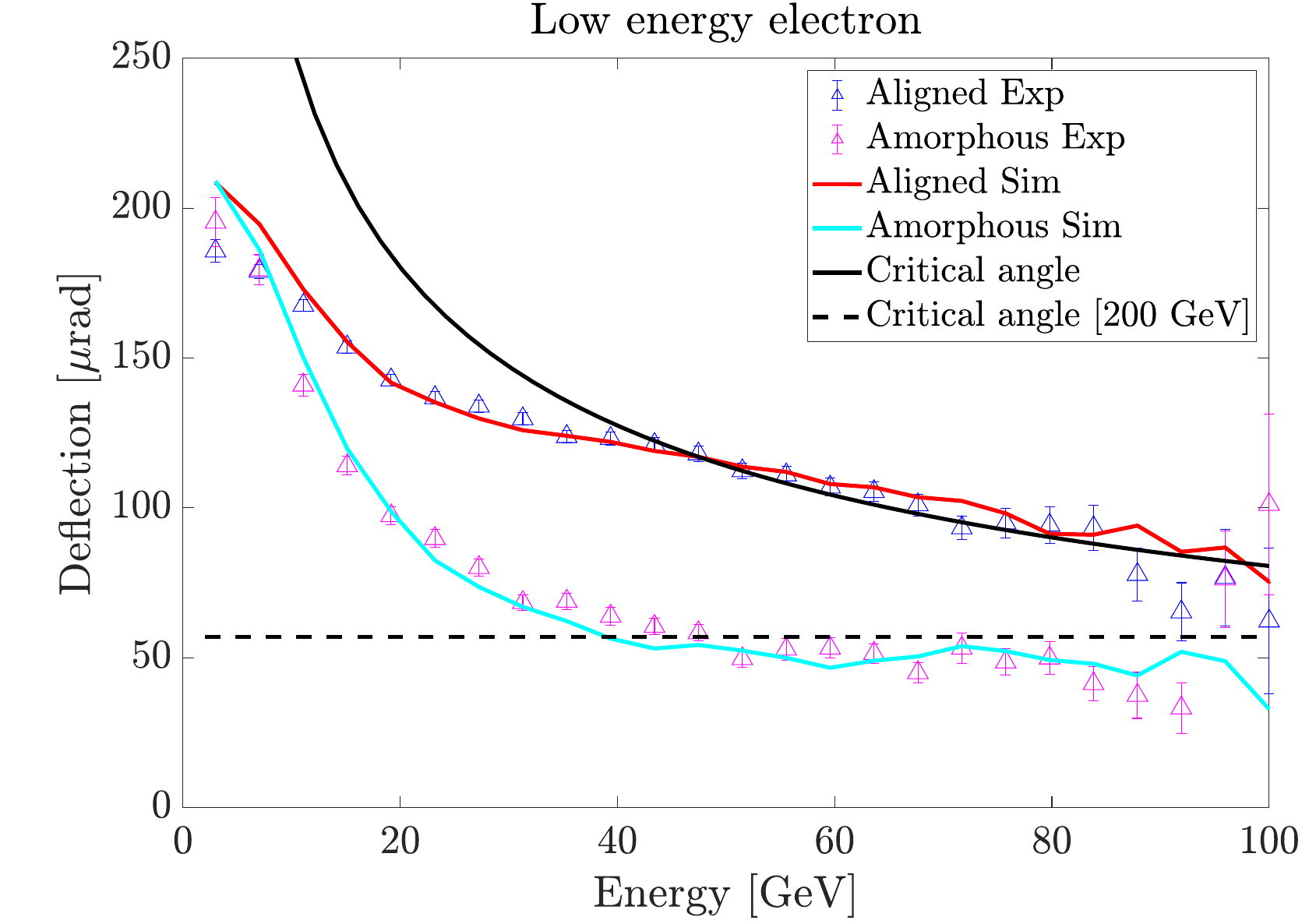} 
\end{subfigure}\\\vspace{30pt}
\begin{subfigure}{\linewidth}\centering
		\includegraphics[width=\linewidth]{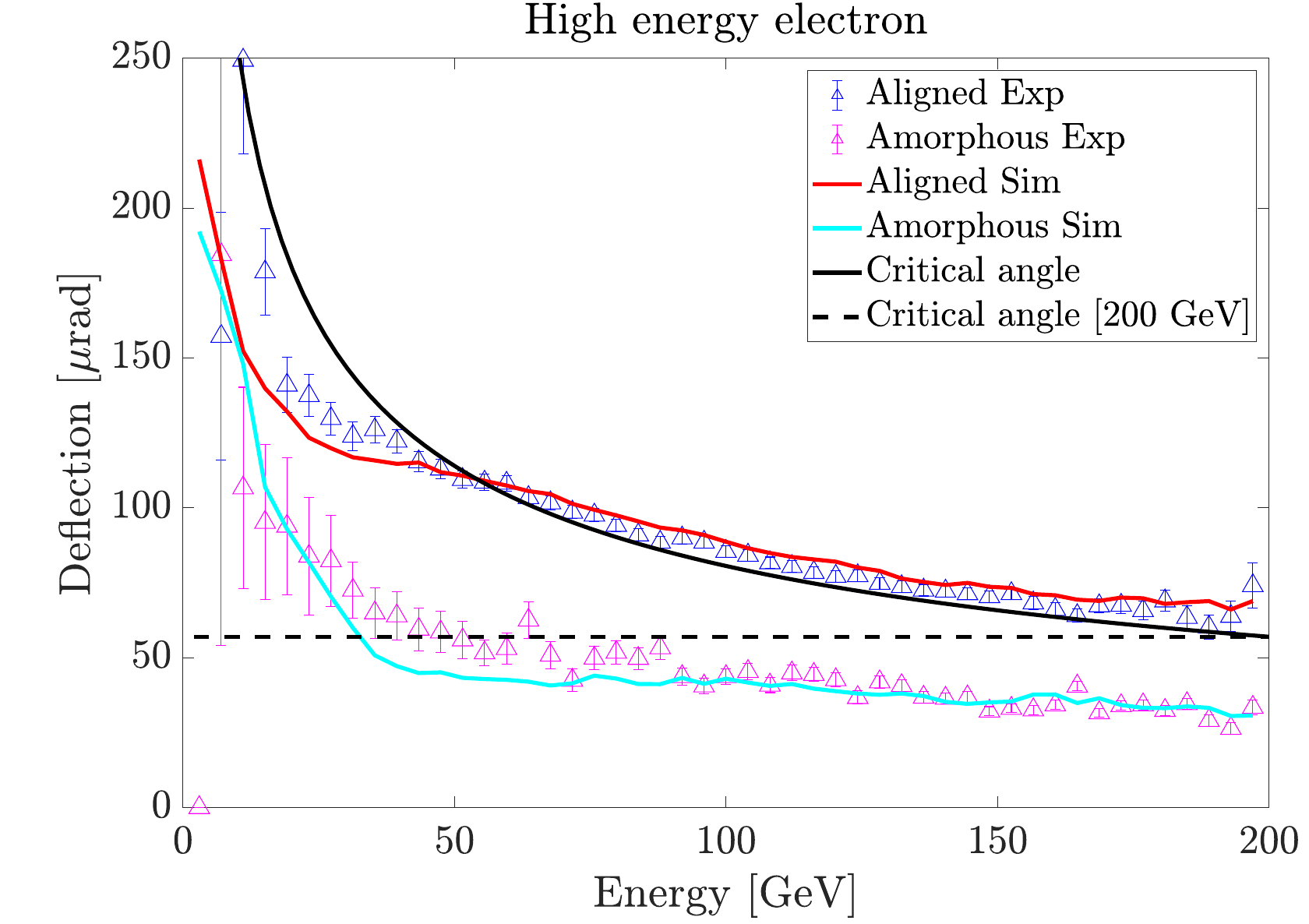} 
\end{subfigure}\vspace{20pt}
\caption{The standard deviation of deflection angle for positrons (top), low energy electrons (middle) and high energy electrons (bottom) for accepted trident events as a function of energy. The black full-drawn line represents the critical angle \cref{eq:psi1} for the final-state particles in each case, while the dotted line represents the critical angle of the primary 200 GeV electron.}
\label{fig:EnergyVsDeflectionCuts}
\end{figure}

In \cref{fig:AlignedSeparation,fig:AmorphSeparation,fig:BackgroundSeparation}, the energy separation ($E_{e+}/(E_{e+}+E_{e-})$) is depicted, again using the lowest energy electron. The energy separation spectra are shown for various photon energies and binned in 22.2 GeV energy bins, making it easier to make direct comparisons between simulation and experiment in a semi differential spectrum. Here the simulated curves are multiplied by the energy efficiency factor $f(E)$ but using the argument $\hbar\omega E_+/(E_++E_-)$ where $\hbar\omega$ is the bin center photon energy pertaining to each sub figure. For low energy photons we see the a clear symmetry around 0.5 where the amorphous and background data keep this symmetry for higher photon energies compared to the aligned data. The asymmetry becomes very pronounced at large photon energies, which is because the ratio uses the low energy electron. On \cref{fig:TridentTypesPhotonEnergy} it is clear that incoherent processes dominate at low energy photon energies, while the coherent processes require a higher photon energy. For photon energies around 100 GeV the maximum $\chi$ value the photon encounters is around $\chi \approx 1.2$, in this region the energy separation is rather localized around 0.5, whereas the incoherent pair production processes have a more flat distribution. This is also visible in the data, as the energy separation for photon energies between 44 to 110 GeV are significantly more flat in the background and amorphous data, compared to the aligned data. For the lowest photon energy bin, the separation is localized around 0.5, which is because the detection efficiency drops significantly for particle below 10 GeV, and a trident with a constituent below this energy is unlikely to be measured. That we recreate this behaviour with the simulation indicates that transverse detector geometry is implemented well in the simulation. On \cref{fig:AlignedSeparation}  curve with and without the coherent direct trident contribution, and see a clear difference for very large photon energies. Because we use the LCFA pair production model in the coherent direct trident process, the energy distribution should be identical between the two simulated curves, but a difference in the photon spectrum could arise.  Since $ \chi = 2.4$ is relatively low, the coherent photon spectrum does not have a large contribution at large photon energies, whereas the virtual photon spectrum extends significantly further for these $\chi$ values. For the 156 GeV to 178 GeV photon energy bin we see a clear influence of the coherent direct trident term. Excluding the process results in almost a factor 2 discrepancy between data and simulation, while a smaller discrepancy is found when including the coherent direct process. At these photon energies, the virtual photon spectrum is extremely dependent on the choice of $b_\mathrm{min}$  \cite{Jackson_b_1975}, which is evident from  \cref{fig:kelnerVsWW,fig:kelnerVsWWPhotonIntegrated}, which might be the cause of the slight discrepancy between data and simulation when excluding the coherent direct trident process in the simulation, for high energy tridents. For large positron energies, the coherent direct process dominates over the incoherent direct process for the present experimental conditions. This is because the pair production process for photon energies above $100$ GeV will experience $\chi_\gamma$ values above 1, which means that the coherent pair production process no longer is exponentially suppressed. The prospects for trident production to be a high intensity positron source for future colliders has been discussed for many years \cite{PhysRevSTAB.10.073501,PhysRevSTAB.17.051003}. Apart from the fact that the number of positrons produced needs to be high, the emittance of the produced positron beam also has to be low to allow for easy injection into the following collider. In \cref{fig:EnergyVsDeflectionCuts} we show the standard deviation of the angle between the incoming and outgoing particle directions in the crystal, the "deflection angle", as a function of energy, for all three constituents of the trident process. Here it is evident that all three particles in a trident event receive a larger transverse momentum for all energies above 20 GeV when produced in an aligned crystal compared to an amorphous crystal. There are several processes at play when penetrating a crystal, either aligned or amorphous, that contribute to an increased transverse momentum. In the amorphous target the dominating processes that contribute to a larger transverse energy is the opening angle between the particles during emission and pair creation, which is of the order $1/\gamma$, and multiple Coulomb scattering on random nuclei throughout the crystal, which also scales as $1/\gamma$. In the aligned crystal the effective continuum electric field and locally varying atomic density plays a major role in this regard as well. In an amorphous crystal, the atomic density is constant which means that there should be no difference between the positron and electron. In an aligned crystal, channeled electrons are confined to an area with a high atomic density, meaning that these electrons are much more likely to scatter incoherently on thermally displaced atoms than a positron, which is repelled by the areas with high atomic density. It was therefore speculated that positrons being produced in such a crystal, would scatter less than in an amorphous crystal, because the positrons on average are located in areas with lower atomic density.

In \cref{fig:EnergyVsDeflectionCuts}, we see the exact opposite behavior for all energies above 10 GeV. The aligned crystal produces positrons with significantly larger transverse momentum than the amorphous crystal. This observation is in part attributed to the moment the pair is created. In the aligned crystal, a pair is most likely formed in an area of strong electric field, where the photon decays. As soon as the pair has been created, the electric field will separate the electron from the positron. This will give them an energy associated with the transverse motion on the order of the potential depth, or an angle corresponding to the critical angle \cref{eq:psi1}, which scales as $1/{\sqrt{\gamma}}$, in opposite directions. After their creation, the only difference between the positron and electron is that the electron undergoes, on average, a greater degree of incoherent scattering on atomic nuclei. In the aligned case, experimental data and simulations show that positrons and the high-energy electrons produced by the trident process follow the critical angle quite closely, \cref{fig:EnergyVsDeflectionCuts}. This implies that the continuum field rather than incoherent scattering dominates their motion. It is also very reassuring to see that our simulations reproduce this simple behavior. Additionally, it should be noted that for the aligned case, the variation in angle of the incident 200 GeV electron during the passage of the crystal corresponds to the critical angle (57 $\mu$rad). Consequently, all products of the trident process will have an angle relative to the incident electron of at least this magnitude. Hence, it is marked in the figures. According to the above discussion, the effect of multiple scattering scales with $1/E$ while the critical angle scales with $1/\sqrt{E}$. This explains the sharp increase in deflection angle for amorphous crystals at lower energies. Due to the experiment's energy cutoff of approximately 10 GeV, deflection angles below this point might be seriously biased, and any structure below this region should be discarded.

\begin{figure}[ht!]
\centering
\begin{subfigure}{1\linewidth}\centering
		\includegraphics[width=\linewidth]{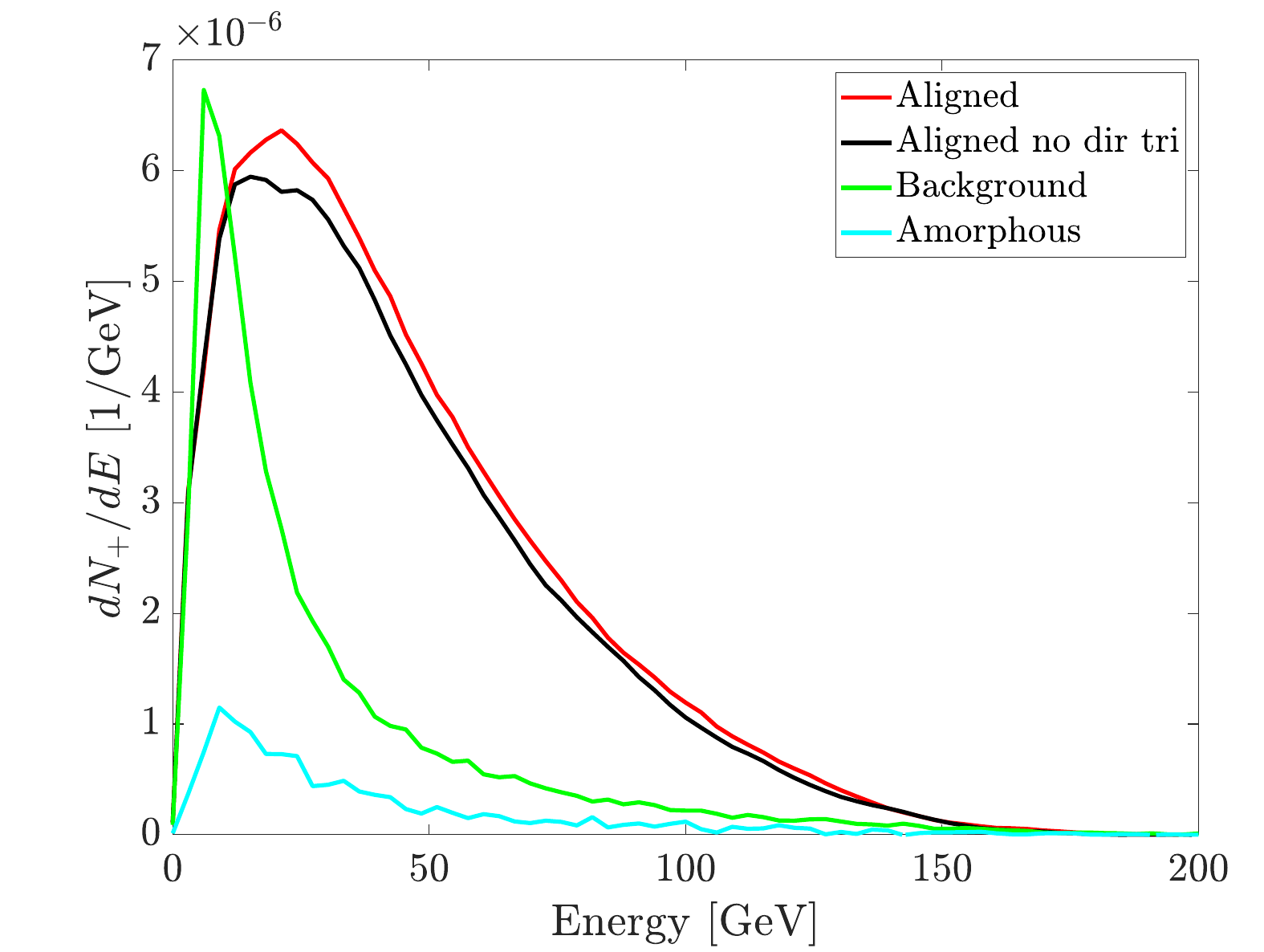} 
\end{subfigure}\\
\begin{subfigure}{1\linewidth}\centering
		\includegraphics[width=\linewidth]{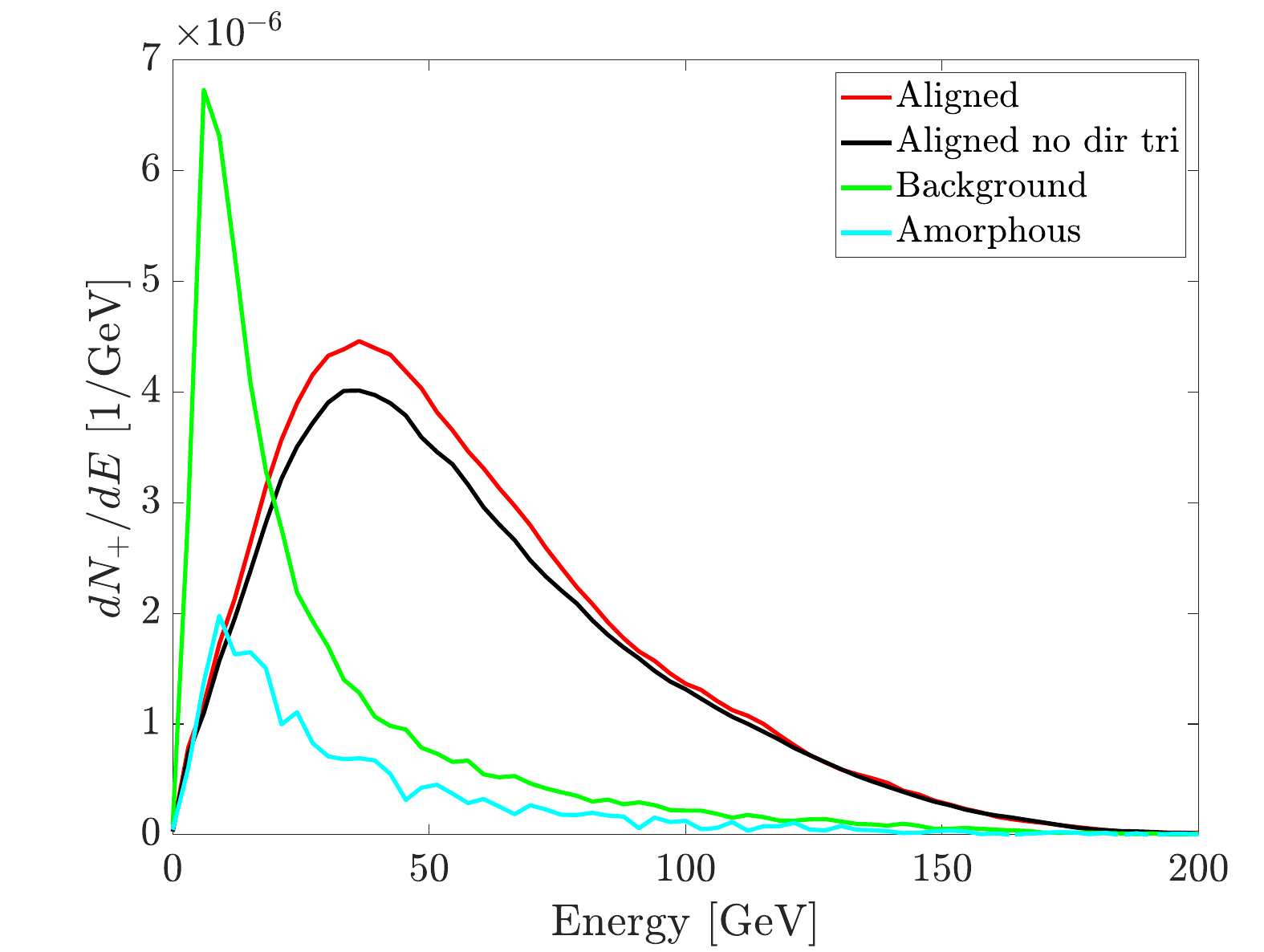} 
\end{subfigure}\\
\caption{Simulated trident spectra for 200 GeV electrons penetrating various crystals. The top figure shows the spectrum for a 50 $\mu$m thick Ge crystal oriented along the $\langle 110 \rangle$ axis and cooled to 100 K. The bottom figure illustrates the spectrum for tungsten 10 $\mu$m thick oriented along the along the $\langle 111 \rangle$ axis. The red curve represents the aligned crystal including all processes. The black curves represent the aligned crystal without the direct coherent process. The background is green while the amorphous crystal is cyan.}
\label{fig:FutureTridentSpectra}
\end{figure}

\section{Future experiments}
In this experiment, we used a crystal with a thickness comparable to its effective radiation length (\cref{eq:effective_rad_length}) when oriented along the axis. 
Therefore, the trident process was dominated by the two-step process, and the direct term had a negligible impact. To measure the effect of the direct coherent trident process, including possible exchange terms, a thin crystal must be used, and $\chi$ must be increased.  In \cref{fig:FutureTridentSpectra}, we present simulated trident spectra for two crystal candidates that may allow us to measure the influence of the direct process realistically:   a germanium crystal of 50 $\mu$m thickness oriented along the $\langle 110 \rangle$ axis while cooled to 100 K and a tungsten crystal of 10 $\mu$m thickness oriented along the $\langle 111 \rangle$ axis. In both cases, a 200 GeV electron is used as a primary particle, and the experimental conditions are the same as those of the present experiment. Germanium crystals are cooled to narrow their atomic string potentials, thereby increasing their maximum field strength. The result is that $\chi$ becomes larger, which in this case reaches $\chi \approx 4$. This change is rather significant since the higher fields allow lower energy photons, real as well as virtual, to convert into pairs in a coherent process rather than in an incoherent process. Since tungsten crystals have a much larger atomic number, a 200 GeV electron reaches $\chi \approx 13$ when oriented along the $\langle 111 \rangle$ axis. There is an interest in this regime because theoretical studies have suggested \cite{KingRuhl2013} that the two-step process is maximally suppressed at $\chi = 10$ by the cross-term between the two-step and one-step terms. In this tungsten configuration, the $\chi$ values are close to the limit of what can be achieved using crystals as a source of strong electric fields, without increasing the initial energy of the electrons. Using the tungsten crystal, we observe the largest effect of the direct process modeled by the WW method. Cooling the germanium crystal experimentally is not an easy process, and it requires considerable engineering to maintain the crystal's axis orientation throughout the cooling process. As a result, using tungsten crystals is easier in this regard, as cooling can be avoided since it has little effect. However, it is challenging to produce these crystals.  Essentially, at 10 $\mu$m, the crystal becomes a foil that blows in the wind and behaves similarly to aluminum foil. The process of producing it is therefore extremely difficult. At 50 $\mu$m a germanium crystal will be structurally solid, and production of germanium crystals has been perfected by the micro-chip industry, making it easy to produce and handle. A further challenge to measuring the direct term is the overall production rate. According to \cref{fig:FutureTridentSpectra}, the rates are more than an order of magnitude lower than in the present experiment shown in \cref{fig:Spectrum}. In the aligned orientation, the background contribution is nearly as large as the signal, whereas the amorphous contribution is several factors lower than the background. This problem could be solved by using more than one particle in each event, as well as increasing the overall beam time. Using the present data analysis algorithm, we were able to reliably analyze events with up to at least 10 primary particles, assuming only one trident was produced per event. Therefore, an attempt to measure the direct term can be made by increasing beam intensity by a factor 10 and beam duration by a factor 2, compared to this experiment. On the other hand, despite the very thin targets -- which in themselves set severe constraints, for example on the permissible background as mentioned above -- the differences between including and excluding the direct trident contribution seen in  \cref{fig:FutureTridentSpectra} is quite small. Moreover, experimentally there is no way of discerning the direct from the sequential trident, and neither the former nor the latter can be experimentally eliminated. It would thus not only be a tough challenge experimentally, but also for the theory part, as the results -- necessarily absolute rather than relative -- must be reliable and precise to the few-percent scale in order to verify the presence of direct tridents.

\section{Conclusion}
This paper provides an even more challenging test of strong field QED for trident events than given in our previous paper \cite{Niel_2023}, in which the integral production of tridents was shown to be very well theoretically described over almost 3 orders of magnitude in yield. Also in the present -- more demanding -- case we can conclude that theory and experimental data are in remarkably good agreement. 
However, all of our strong-field results for trident production are dominated by the two-step process. The direct trident process has yet to be measured and tested by an experiment, and we have discussed some of the technical challenges involved in such a task.

\section{Acknowledgments}
We acknowledge the expertise of Frank Daugaard and Erik Loft Larsen in setting up the electronics and logic circuits, and for their generous help with installing the detectors, vacuum chambers and helium-pipes.\\ The numerical results presented in this work were partly obtained at the Centre for Scientific Computing Aarhus (CSCAA) and with support from Nvidia's GPU grant program. This work was partially supported by the U.S. National Science Foundation (Grant No. PHY-1535696, and PHY-2012549) and from the Danish National Instrument Center for CERN Experiments (NICE), \url{https://nice.ku.dk/}.

\bibliography{references}

\newpage

\appendix

\section{Chebyshev implementations}\label{sec:AppendixA}
In this appendix we show the specific implementations of the Cheyshev series mention in \cref{sec:EMProcesses} and \cref{sec:AMProcesses}.

\subsection{Coherent Direct Trident} 
The integral in \cref{eq:TridentProbability} is evaluated as a function of $\chi$ and represented as
\begin{equation}\label{TridentProductionProbCheby1}
             \frac{\text{d}P_{\text{WW}}^{\text{trident}}}{\text{d}t} =R_1(t_{\chi})\frac{2\alpha^2 m^2}{\pi E}\text{e}^{-2/\chi}, \quad t_{\chi} = 2\frac{\chi-0.1}{3-0.1}-1, 
\end{equation}
when $0.1<\chi <3$ and by
\begin{equation}\label{TridentProductionProbCheby2}
             \frac{\text{d}P_{\text{WW}}^{\text{trident}}}{\text{d}t} =R_2(t_{\chi})\frac{2\alpha^2 m^2}{\pi E}, \quad t_{\chi} = 2\frac{\chi-3}{1000-3}-1, 
\end{equation}
when $3<\chi <1000$.
The functions $R(t)$ are then the fitted Chebyshev series which in our case consists of 40 and 50 terms respectively. The functions are in practice only evaluated for $\chi > 0.2$ to avoid evaluating the probabilities in locations where the field is small and the resulting probability is negligible.

The function \cref{PropabiblityDensityTridentPhoton} is inverted to express $x$ as a function of the random number $r$ and $\chi$ which we can fit with a 2-dimensional Chebyshev series $R(t_r,t_{\chi})$. For better agreement between fit and function we make three separate Chebyshev series. The first series is defined by
\begin{equation}\label{RatioTridentPhoton1}
    x = \frac{R_1(t_r,t_{\chi})}{r\sqrt{\chi}}, 
\end{equation}
with 
\begin{equation}
    t_{\chi} = 2\frac{\chi-0.2}{1-0.2}-1, \quad t_{r} = 2\frac{r}{0.03}-1
\end{equation}
when $0 < r < 0.03$ and $0.2 < \chi < 1$. Second fit is defined by
\begin{equation}\label{RatioTridentPhoton2}
    x = \frac{R_2(t_r,t_{\chi})}{r}, 
\end{equation}
with 
\begin{equation}
    t_{\chi} = 2\frac{\chi-0.2}{1-0.2}-1, \quad t_{r} = 2\frac{r-0.03}{1-0.03}-1
\end{equation}
when $0.03 < r < 1$ and $1 < \chi < 50$. The last fit is defined by the same function as in \cref{RatioTridentPhoton2} 
but with 
\begin{equation}
    t_{\chi} = 2\frac{\chi-1}{50-1}-1, \quad t_{r} = 2r-1
\end{equation}
when $0 < r < 1$ and $1 < \chi < 50$.
Both of these series are fitted using $65\times65$ parameters.

The function \cref{PropabiblityDensityTridentPair} is inverted to express $y$ as a function of $x$, $\chi$ and $r$, leaving us with a three dimensional Chebyshev series, which is defined through:
\begin{equation}\label{RatioTridentPair}
    y = R(t_r,t_{\chi},t_x).
\end{equation}
We again fit three separate series in the various regions of $x$, $\chi$ and $r$, but all according to \cref{RatioTridentPair}.
The first region is defined as
\begin{equation}
    t_{\chi} = 2\frac{\chi-0.2}{2-0.2}-1, \quad t_{r} = 2\frac{r}{0.03}-1 , \quad t_{x} = 2x-1,
\end{equation}
when $0 < r < 0.03$, $0.2 < \chi < 2$ and $0 < x < 1$. The second region is defined by
\begin{equation}
    t_{\chi} = 2\frac{\chi-0.2}{2-0.2}-1, \quad t_{r} = 2\frac{r-0.03}{0.5-0.03}-1 , \quad t_{x} = 2x-1,
\end{equation}
when $0.03 < r < 0.5$, $0.02 < \chi < 2$ and $0<x<1$. The last region is defined by
\begin{equation}
    t_{\chi} = 2\frac{\chi-2}{50-2}-1, \quad t_{r} = 2\frac{r}{0.5}-1 , \quad t_{x} = 2x-1,
\end{equation}
when $0 < r < 0.5$, $2 < \chi < 50$ and $0<x<1$. Since the pair spectrum is symmetric in $y$ around $r = 0.5$, we only evaluate the spectrum in the region $0<r<0.5$. A second uniformly distributed random number is drawn, $0<r_2<1$, where we use the previously found $y$ value if $r_2 < 0.5$ and use $y' = x-y$ if $r_2 > 0.5$. 
All three series are fitted using $25\times 25 \times 25$ parameters.

\subsection{Bremsstrahlung}
The inverse of \cref{eq:BremsstrahlungInversePhoton} is evaluated as a function of $E$ and $r$, and a Chebyshev series is fitted directly to the function.
\begin{equation}
    E_\gamma = R(t_r, t_E),
\end{equation}
in two energy regions. The first series is defined by 
\begin{equation}
    t_{E} = 2\frac{E-1}{10^3-1}-1, \quad t_{r} = 2r-1
\end{equation}
when $0 < r < 1$ and $1 < E < 10^3$. Second fit is defined by
\begin{equation}
    t_{E} = 2\frac{E-10^3}{10^6-10^3}-1, \quad t_{r} = 2r-1
\end{equation}
when $0 < r < 1$ and $10^3 < E < 10^6$. Here the energy $E$ is in units of MeV. The two fits are therefore defined in the region between 1 MeV and 1 TeV. 
Both series are fitted using $60 \times 60$ parameters.

\subsection{Incoherent Direct Trident}
The probability per unit time is evaluated by integrating \cref{eq:KelnerTrident} as in \cref{eq:TridentProbability}, as a function of the charge number $Z$, and fitted with a Chebyshev series for quick evaluation during each timestep. A single series with 30 parameters is fitted in the region $1 < Z < 200$, and is defined by
\begin{equation}
                 \frac{\text{d}P_\text{Kel}}{\text{d}t} =R(t_{Z})\frac{2NZ^2\alpha^4}{\pi m^2}, \quad t_{Z} = 2\frac{Z-1}{200-1}-1, 
\end{equation}
where $R(t_Z)$ is the Chebyshev series. 

If an incoherent trident is produced, a value for $x$ and $y$ is to be found.
We define the cumulative probability density function and set it equal to a random number times the total probability:
\begin{equation}
\label{PropabiblityDensityKelnerPhoton}
r\, \frac{\text{d}P_{\text{Kel}}}{\text{d}t}
=
 \int^{x}_0\int^{x'}_0\frac{\text{d}P_\text{Kel}}{\text{d}x'\text{d}y\text{d}t} \text{d}y\text{d}x'.
\end{equation}
We invert the equation and solve for the ratio $x$. In this way we can express $x$ as a function of the random number $r$ and $Z$. This function can be fitted with good agreement using a single series with parameters in the $r$ dimension, 20 parameters in the $Z$ dimension and is defined by
\begin{equation}\label{RatioKelnerPhoton1}
    x = R(t_r,t_{Z}), 
\end{equation}
with 
\begin{equation}
    t_{Z} = 2\frac{Z-1}{200-1}-1, \quad t_{r} = 2r-1
\end{equation}
when $0 < r < 1$ and $1 < Z < 200$. 

After picking a value for $x$ we find the ratio $y$.  
The cumulative probability density function is found and set equal to a random number times the total probability for a specific $x$:
\begin{equation}
\label{PropabiblityDensityTridentPhoton}
r\, \frac{\text{d}P_{\text{Kel}}}{\text{d}x\text{d}t}(x)
=
 \int^{y}_0 \frac{\text{d}P_\text{Kel}}{\text{d}x\text{d}y'\text{d}t} \text{d}y'.
\end{equation}
This function is inverted and solved for $y$ as a function of $x$, $Z$ and $r$, leaving us with a three dimensional Chebyshev series with 25 parameters in the $x$ dimension, 10 parameters in the $Z$ dimension and $25$ parameters in the $r$ dimension. The series is fit directly to the function
\begin{equation}\label{RatioTridentPair}
    y = R(t_r,t_{Z},t_x),
\end{equation}
with
\begin{equation}
    t_{Z} = 2\frac{Z-1}{200-1}-1, \quad t_{r} = 2\frac{r}{0.5}-1 , \quad t_{x} = 2x-1,
\end{equation}
when $0 < r < 0.5$, $1 < Z < 200$ and $0<x<1$. Since the pair spectrum is asymmetric for $y$ around $r = 0.5$, we only evaluate the spectrum in the region $0<r<0.5$. A second uniformly distributed random number is drawn, $0<r_2<1$, where we use the previously found $y$ value if $r_2 < 0.5$ and use $y' = x-y$ if $r_2 > 0.5$. 
We are able to use significantly less fitting parameters in the $Z$ dimension because the curves along this dimension vary slowly compared to the LCFA model which is extremely sensitive to $\chi$ in the region around $\chi = 1$.

\hspace*{1cm}
\section{Kelner's $\Phi_A$ and $\Phi_B$}\label{sec:AppendixB}
Here we provide the expressions for the two lengthy quantities used in \cref{eq:KelnerTrident}.

The first, $\Phi_A$, is defined by
\begin{multline}
        \Phi_A = 2\ln\left(183Z^{-1/3}\sqrt{1+\xi}\right)\left[a_1\ln\left(1+\frac{1}{\xi}\right)-b_1\right.\\\left.-\frac{c_1}{1+\xi}\right]+a_1S(1-\frac{1}{1+\xi})-d_1\ln\left(1+\frac{1}{\xi}\right)- \frac{2c_1}{3(1+\xi)} + \frac{2}{9} \beta \zeta,
\end{multline}
with the Spence function $S(z)$ defined as
\begin{equation}
    S(z) = \int^z_1\frac{\ln(t)}{1-t}\text{d}t
\end{equation}
together with the remaining parameters
\begin{gather}
    a_1 = \left(\beta^2 + \zeta^2 + \frac{2}{3}\beta\zeta\right)\left(1+\frac{x}{2(\frac{1}{x}-1)}
    \right)+\frac{4}{3}\xi(1-\zeta\beta)\\
    b_1 = \zeta^2 + \beta^2 + \frac{2}{3}\zeta\beta\\
    c_1 = \frac{1}{3}\xi+\frac{1}{3}(\beta - \zeta)^2+\frac{x}{(1/x-1)}\\
    d_1 = b_1\xi+\frac{1}{9}\beta\zeta\left( \frac{1}{1-x}+1-x\right)+\frac{1}{9}\xi(1+2\zeta\beta).
\end{gather}
Kelner's $\Phi_B$ is likewise defined as
\begin{multline}
        \Phi_B = 2\ln\left(183Z^{-1/3}\sqrt{1+\frac{1}{\xi}}\right)\left[a_2\ln\left(1+\xi\right)+b_2\beta\zeta\phantom{\frac{1}{2}}\right.\\\left.+\frac{c_2\xi}{1+\xi}\right]+a_2S(1-\frac{\xi}{1+\xi})+d_2\ln\left(1+\xi\right)+ \frac{2c_2\xi}{3(1+\xi)} + \frac{2}{9} \beta \zeta,
\end{multline}
together with the parameters
\begin{gather}
    a_2 = \left(\frac{\beta^2}{2} +\frac{\zeta^2}{2}-\frac{\beta\zeta}{\xi}\right)\left(\frac{1}{3}+\frac{1}{1-x}-x\right) - \frac{1}{3}\\
    b_2 = \left(\frac{1}{3}+\frac{1}{1-x}-x\right)\\
    c_2 = \frac{4}{3}\beta\zeta-\frac{x}{6(\frac{1}{x}-1)}(\zeta^2+\beta^2) + \frac{1}{3}\\
    d_2 = b_2\frac{\zeta\beta}{\xi}-\frac{2}{9}\left( \frac{\zeta\beta}{\xi}-\frac{\beta^2}{2} -\frac{\zeta^2}{2} \right) +\frac{1}{9}.
\end{gather}

\end{document}